\documentclass[onecolumn, draftcls, 12pt]{IEEEtran}

\usepackage{amsfonts}
\usepackage{mathrsfs}
\usepackage{amssymb,amsmath}

\usepackage{algorithm}
\usepackage{algorithmic}
\usepackage{amsmath,amssymb,epsfig,graphics,subfigure}
\usepackage{theorem}
\usepackage{array,color}
\usepackage[compress]{cite}

\theoremheaderfont{\normalfont\bfseries}

\begin{document}

\title{Matrix-Monotonic Optimization for MIMO Systems}

\author{Chengwen Xing, Shaodan Ma and Yiqing Zhou
\thanks{This work was in part supported by 111 Project of China under Grant. B14010, National Natural Science
Foundation of China (NSFC) under Grant No. 61101130 and the Project of China Mobile Research Institute under Grant No. [2014]451. }

\thanks{C. Xing is with the School of Information and Electronics, Beijing Institute of Technology, Beijing 100081, China. Phone : (86)
010 6891 5838, (e-mail: chengwenxing@ieee.org). }
\thanks{S. Ma is with the Department of Electrical and Computer Engineering, University of Macau, Macao (e-mail:
shaodanma@umac.mo).}

\thanks{Y. Zhou is with the Wireless Research Center, Institute of Computing
Technology, Chinese Academy of Sciences, Beijing 100190, China (e-mail:
zhouyiqing@ict.ac.cn).}}

\maketitle

\begin{abstract}
For MIMO systems, due to the deployment of multiple antennas at both the transmitter and the receiver, the design variables e.g., precoders, equalizers, training sequences, etc. are usually matrices. It is well known that matrix operations are usually more complicated compared to their vector counterparts. In order to overcome the high complexity resulting from matrix variables, in this paper we investigate a class of elegant multi-objective optimization problems, namely matrix-monotonic optimization problems (MMOPs). In our work, various representative MIMO optimization problems are unified into a framework of matrix-monotonic optimization, which includes linear transceiver design, nonlinear transceiver design, training sequence design, radar waveform optimization, the corresponding robust design and so on as its special cases. Then exploiting the framework of matrix-monotonic optimization the optimal structures of the considered matrix variables can be derived first. Based on the optimal structure, the matrix-variate optimization problems can be greatly simplified into the ones with only vector variables. In particular, the dimension of the new vector variable is equal to the minimum number of columns and rows of the original matrix variable. Finally, we also extend our work to some more general cases with multiple matrix variables.
\end{abstract}

\newpage
\section{Introduction}
\label{sect:intro}

 Multiple-input multiple-output (MIMO) technology is a great success in the research of wireless communication theory. Since Telatar's landmark paper \cite{Telatar1999}, more than ten years have passed and MIMO technology has been well studied and developed. Many elegant and important results have been derived and published. Now it is well-known for wireless engineers and researchers that deployment of multiple antennas at transmitter or receiver or both can greatly improve wireless system performance. As a result, in order to gain the virtues promised by MIMO technology continuously, multiple-antenna array has become one of the most essential parts of a lot of communication systems such as cognitive radio, cooperative communications, even radar systems and so on.

 Different from  single-antenna wireless communication systems, the optimization problems for MIMO systems usually have much higher dimensions, e.g., more parameters to be estimated and more variables to be optimized. In general, MIMO designs are formulated as multivariate optimization problems with high complexity and high dimension. To reduce the complexity, it is necessary to transform high-dimensional optimization problems with matrix variables into low-dimensional ones with only vector variables. A common approach for this transformation is to find the structure information of optimal matrix variables which then enables the reduction of the dimension and number of variables. The derivation of optimal structures usually the diagonalizable stuctures of the matrix variables thus becomes the core of many wireless designs including transceiver design \cite{Palomar03,Kotecha2004,Rong2009TWC,Guan08,Tang07,Medina07,Shenouda2008,Amico2008}, training sequence design \cite{FFGao08,Biguesh2006,Liu2007,Katselis2008}, radar waveform optimization \cite{Tang2010}, their corresponding robust design \cite{Ding09,Ding10,Zhang08,Xing10,XingICSPCC} and so on.


 In the existing literature, the common logic to derive the optimal structures relies on optimization theory. Generally speaking, there are two kinds of methods to derive the optimal diagonalizable structures, i.e., Karush-Kuhn-Tucker (KKT) condition based methods \cite{Sampth01,Ding09,Ding10,Schizas07} and majorization theory based methods \cite{Palomar03,Rong2009TWC,Shenouda2008,Amico2008}. The KKT condition based methods usually exploit the fact that with certain regularity conditions satisfied, KKT conditions are the necessary conditions for optimal solutions. Then if all the solutions of matrix variables satisfying KKT conditions have a common structure, this structure is definitely the optimal structure of the optimal solutions. For this kind of methods, the main problem is when the problem is complicated, the complicated formulas of KKT conditions will prohibit us from deriving the structures. On the other hand, the idea behind majorization theory based methods is to transform various performance metrics as the functions of the diagonal elements of mean-square-error (MSE) matrix. In this case, another problem arises that as for different systems, MSE matrices are totally different and even the derivation logics will be different from case to case \cite{Rong2009TWC}. Thus the majorization theory based method is system dependent as its application is limited to MSE matrix formulations that are functions of system parameters.

 With the development of wireless systems, wireless designs are faced with more and more challenges. One of the challenges comes from channel estimation errors. To mitigate the negative effects of channel estimation errors, it is necessary to take them into account in the designs and thus robust designs become more and more important \cite{JHWang2009,HMWang2012,WXu2012,Xing1012}\footnote{As there is a rich body of papers on this topic, we only list the most relevant work here.}. Faced with these ever increasing challenges, the existing technologies are insufficient. Recently, similar to convexity, the monotonicity of objective functions has been accepted as a valuable property that can be exploited to optimize wireless systems \cite{Qian2010,Wolfgang2012,Larsson2010}. As revealed in the literature, efficient and proper utilization of monotonicity is of great importance for resource allocation and interference coordination. Moreover, as shown in  \cite{Qian2010,Larsson2010}, several optimization problems that are even not convex can also be efficiently solved using monotonicity property. Since monotonicity is as fundamental as convexity, it can facilitate numerical computation of the optimal solutions \cite{Qian2010,Wolfgang2012,Larsson2010}. However, whether this monotonicity property can be exploited to derive the optimal structures of matrix variables is unclear yet and will be investigated in this paper.

In this paper, we investigates a series of matrix-variate optimization problems (MVOPs) from the perspective of monotonicity in positive semi-definite matrices. By exploiting an elegant and powerful framework of matrix-monotonic optimization, the optimal structures of matrix variables can be derived in a unified manner. Specifically, the optimization problems which can be unified into our framework include the following wireless designs.

\noindent $\bullet$ Robust capacity maximization transceiver design for point-to-point MIMO systems.

\noindent $\bullet$ Training sequence design for point-to-point MIMO systems based on mutual information maximization criterion. This problem is closely related to radar waveform optimization based on mutual information maximization.

\noindent $\bullet$ Robust linear minimum mean square error (LMMSE) transceiver design for point-to-point MIMO systems.

\noindent $\bullet$ Training sequence design for point-to-point MIMO systems based on minimum mean square error (MMSE) criterion.  This design is also closely related to MMSE radar waveform optimization.

\noindent $\bullet$ Robust LMMSE transceiver design for dual-hop amplify-and-forward (AF) MIMO relaying systems without source processing. This problem is closely related to the problem of signal detection in sensor networks.

\noindent $\bullet$ Robust capacity maximization transceiver design for dual-hop AF MIMO relaying systems without source processing. This problem is also closely related to the problem of signal detection in sensor networks.

\noindent $\bullet$ Robust linear transceiver designs for point-to-point MIMO systems with additive Shur-convex/concave objective functions.

\noindent $\bullet$ Robust nonlinear transceiver designs for point-to-point MIMO systems with decision feedback equalizer (DFE) receiver or Tomlinson-Harashima precoding (THP) transmitter with multiplicative Shur-convex/concave objective functions.

\noindent $\bullet$ Robust linear transceiver designs for multi-hop AF MIMO relaying systems with additive Shur-convex/concave objective functions.

\noindent $\bullet$ Robust nonlinear transceiver designs for multi-hop AF MIMO relaying systems with DFE receiver or THP transmitter with multiplicative Shur-convex/concave objective functions.

\noindent $\bullet$ Transceiver designs for multicarrier MIMO systems.

At the beginning we would like to point out that our work is  significantly different from the relevant and important work \cite{Jorswieck07}, as in our work the objective functions of the considered matrix-monotonic optimization problems (MMOPs) are Hermitian matrices instead of a scalar function of the covariance matrices of transmit signals. In other words, we restrict our attention to matrix version multi-objective optimization problems \cite[P. 177]{Boyd04}. Furthermore, different from the existing works  \cite{Qian2010,Wolfgang2012,Larsson2010}, our work focuses on the derivation of the structure of the optimal solutions of MVOP in closed form instead of numerically computing the optimal solutions. Based on the derived optimal structures, the considered MVOPs will be greatly simplified to low-dimensional problems with only vector variables.

Our work can be regarded as a complementary work to \cite{Palomar2006} and \cite{Jorswieck07} and we expect the proposed work can have more applications in the future. The main contributions of our work are summaried as follows.

\noindent $\bullet$ Firstly, it is shown that for MIMO systems various optimization problems with matrix variables can be transformed to MMOPs. Exploiting the matrix monotonicity in positive semi-definite matrices, a framework of matrix-monotonic optimization is built for the derivation of optimal diagonalizable structure. This framework also reveals the relationships among different MVOPs.

\noindent $\bullet$ Secondly, with the optimal structure, the set of Pareto optimal solutions is also analyzed based on multi-objective optimization theory.

\noindent $\bullet$ Thirdly, the unitary-matrix-based transformations from the considered MVOPs to MMOPs are also investigated in detail. Several important matrix inequalities are given as the theoretical basis.

\noindent $\bullet$ Finally, we take a step further to investigate more general MMOPs with more than one matrix variate. These cases are closely related with multi-hop amplify-and-forward MIMO relaying networks and multi-carrier MIMO systems.

The rest of this paper is organized as follows. In Section~\ref{Sect: Problem_Unification}, various wireless designs for MIMO systems are discussed case by case and then they are unified into a single optimization problem which can be solve relying on the framework of MMOP. After that, In Section~\ref{Sect: Fundamentals} the fundamentals of MMOP are investigated in details and the corresponding unitary matrix based transforms to MMOPs are elaborated in Section~\ref{Sect: Unitary_Transform}. Then in Section~\ref{Sect: Multiple_Matrix_Variable} the results are extended to some more general cases with multiple matrix variables. The performance of several most representative cases is assessed by simulations in Section~\ref{Sect: Simulations}. Finally, the conclusions are drawn in Section~\ref{Sect: Conclusions}.

\noindent \textbf{Notation:} The following notations are used throughout this paper. Boldface
lowercase letters denote vectors, while boldface uppercase letters
denote matrices. The notation ${\bf{Z}}^{\rm{H}}$ denotes the
Hermitian transpose of the matrix ${\bf{Z}}$, and ${\rm{Tr}}({\bf{Z}})$ is the
trace of the matrix ${\bf{Z}}$. The symbol ${\bf{I}}_{M}$ denotes an
$M \times M$ identity matrix, while ${\bf{0}}_{M,N}$ denotes an $M
\times N$ all-zero matrix. The notation ${\bf{Z}}^{1/2}$ is the
Hermitian square root of the positive semi-definite matrix
${\bf{Z}}$, such that ${\bf{Z}}^{1/2}{\bf{Z}}^{1/2}={\bf{Z}}$ and
${\bf{Z}}^{1/2}$ is also a Hermitian matrix. For a rectangular diagonal matrix ${\boldsymbol \Lambda}$, ${\boldsymbol \Lambda} \searrow$ denotes the main diagonal elements in decreasing order and ${\boldsymbol \Lambda} \nearrow$ denotes the main diagonal elements in increasing order. For two Hermitian matrices, ${\bf{C}} \succeq
{\bf{D}}$ means that ${\bf{C}}-{\bf{D}}$ is a positive semi-definite
matrix.

\section{Problem Unification}
\label{Sect: Problem_Unification}

In this paper we focus on a special class of matrix-variate optimization  problems (MVOPs) for MIMO systems. The considered MVOPs can be transformed into an elegant and unified  multi-objective problem via some unitary-matrix-based transformations. The resulting multi-objective problem is termed matrix-monotonic optimization problem (MMOP) as the matrix monotonicity of its objective function can be exploited to greatly simplify the optimal solution derivation. The structure of the optimal solutions can be clearly derived first. As a result the remaining unknown variable is simplified from a matrix to a vector. In most cases, the optimal structures also reveal the physical meanings of the design variables. In the following, we list the considered MVOPs case by case. It should be noticed that in MIMO systems training sequence designs and precoder designs are closely related with each other. Roughly speaking there is a duality between them \cite{Vergara2006}. This is the reason why these two kinds of optimization problems can be discussed from a unified viewpoint.

\noindent {\textbf{(a) Mutual Information Maximization:}}

For wireless designs, mutual information maximization is a natural choice for the transceiver design as it reflects how much information can be transmitted over wireless channels. Note that mutual information can also act as a metric reflecting how much the transmitted and received signals are related with each other. Therefore this metric can also be exploited to design training sequences or pilots. With the objective of mutual information maximization, a general optimization problem for MIMO systems is formulated as
\begin{align}
\label{case_1}
{\textbf{Case 1:}} \ \ & \min_{{\bf{X}}} \ \
-{\rm{log}}|{\bf{X}}^{\rm{H}}{\bf{H}}^{\rm{H}}{\bf{K}}_{\bf{X}}^{-1}{\bf{H}}{\bf{X}}
+{\bf{N}}|  \nonumber \\
& \ {\rm{s.t.}} \ \ \  {\bf{K}}_{\bf{X}}={\rm{Tr}}({\bf{X}}{\bf{X}}^{\rm{H}}{\boldsymbol \Psi}){\boldsymbol \Sigma}+{\sigma}_{n}^2{\bf{I}}  \nonumber \\
& \ \ \ \ \ \ \ \ {\rm{Tr}}({\bf{X}}{\bf{X}}^{\rm{H}}) \le P,
\end{align}where ${\bf{X}}$ is an unknown $ N_T\times N$ complex matrix variate. On the other hand, ${\bf{H}}$, ${\bf{N}}$, ${\boldsymbol \Psi}$ and ${\boldsymbol \Sigma}$ are given constant matrices. Furthermore, ${\bf{N}}$, ${\boldsymbol \Psi}$ and ${\boldsymbol \Sigma}$ are all positive semi-definite. When  ${\boldsymbol \Psi}={\bf{0}}$ and ${\bf{H}}={\bf{I}}$, (\ref{case_1}) reduces to training design for MIMO systems \cite{Biguesh2006}. The case of ${\boldsymbol \Psi}={\bf{0}}$ and ${\bf{N}}={\bf{I}}$ corresponds to the linear transceiver design with perfect CSI \cite{Telatar1999}. For robust linear transceiver designs, the matrices ${\boldsymbol \Psi}$ and  ${\boldsymbol \Sigma}$ are functions of training sequences and channel estimators \cite{Ding09}. The detailed discussion of the formulations of ${\boldsymbol \Psi}$ and  ${\boldsymbol \Sigma}$ is out of the scope of this paper. The interested readers are referred to \cite{Ding09,Xing10} and the references therein. In our work, the matrix $\sigma_n^2{\bf{I}}$ always represents the covariance matrix of noise signals and the symbol $P$ in the constraint denotes the maximum power limit.

It should be noticed that although several MIMO designs are almost the same in nature, the specific mathematical formulations may be different.
For the sake of completeness, an optimization problem closely relevant to \textbf{Case 1} is also given in the following
\begin{align}
\label{case_2}
{\textbf{Case 2:}} \ \ & \min_{{\bf{X}}} \ \
-{\rm{log}}|{\bf{A}}^{\rm{H}}{\bf{X}}^{\rm{H}}{\bf{H}}^
{\rm{H}}{\bf{K}}_{\bf{X}}^{-1}{\bf{H}}{\bf{X}}{\bf{A}}
+{\bf{I}}|  \nonumber \\
& \ {\rm{s.t.}} \ \ \ {\bf{K}}_{\bf{X}}={\rm{Tr}}({\bf{X}}{\bf{X}}^{\rm{H}}{\boldsymbol \Psi}){\boldsymbol \Sigma}+{\sigma}_{n}^2{\bf{I}}  \nonumber \\
& \ \ \ \ \ \ \ \ {\rm{Tr}}({\bf{X}}{\bf{X}}^{\rm{H}}) \le P,
\end{align}where ${\bf{A}}$ is a constant complex matrix. Notice that when both ${\bf{N}}$ and ${\bf{A}}$ have full rank, the optimization problem (\ref{case_2}) is in nature the same as (\ref{case_1}), but they have different mathematical formulations.

\noindent {\textbf{(b) MSE Minimization:}}

In contrast to mutual information maximization, mean-square-error (MSE) minimization is another most important performance metric, which indicates how accurately a signal can be recovered from the noise corrupted observations. It is also widely used for training designs and transceiver designs. For MIMO systems, the optimization problem of MSE minimization can be written as
\begin{align}
\label{case_3}
{\textbf{Case 3:}} \ \ & \min_{{\bf{X}}} \ \
{\rm{Tr}}[({\bf{X}}^{\rm{H}}{\bf{H}}^
{\rm{H}}{\bf{K}}_{\bf{X}}^{-1}{\bf{H}}{\bf{X}}
+{\bf{N}})^{-1}]  \nonumber \\
& \ {\rm{s.t.}} \ \ \ {\bf{K}}_{\bf{X}}={\rm{Tr}}({\bf{X}}{\bf{X}}^{\rm{H}}{\boldsymbol \Psi}){\boldsymbol \Sigma}+{\sigma}_{n}^2{\bf{I}}  \nonumber \\
& \ \ \ \ \ \ \ \ {\rm{Tr}}({\bf{X}}{\bf{X}}^{\rm{H}}) \le P.
\end{align} When ${\boldsymbol \Psi}={\bf{0}}$ and ${\bf{N}}={\bf{I}}$, the optimization problem (\ref{case_3}) reduces to the transceiver design with perfect CSI \cite{Palomar03}. For robust transceiver designs, ${\boldsymbol \Psi}$ and  ${\boldsymbol \Sigma}$ are also functions of training sequences and channel estimators \cite{Ding10}. In addition, when  ${\boldsymbol \Psi}={\bf{0}}$ and ${\bf{H}}={\bf{I}}$, (\ref{case_3}) will reduce to the training sequence designs for MIMO systems \cite{Katselis2008}.

For training designs, the problem formulation usually involves Kronecker products which come from the matrix variable position exchanges. Then the optimization problem (\ref{case_3}) has an accompanying optimization problem with the following formula
\begin{align}
\label{case_4}
{\textbf{Case 4:}} \ \ & \min_{{\bf{X}}} \ \
{\rm{Tr}}[\left(({\bf{X}}^{\rm{H}}{\bf{H}}^
{\rm{H}}{\bf{K}}_{\bf{X}}^{-1}{\bf{H}}{\bf{X}})\otimes{\bf{M}}
+{\bf{N}}\otimes {\bf{M}}\right)^{-1}]  \nonumber \\
& \ {\rm{s.t.}} \ \ \ {\bf{K}}_{\bf{X}}={\rm{Tr}}({\bf{X}}{\bf{X}}^{\rm{H}}{\boldsymbol \Psi}){\boldsymbol \Sigma}+{\sigma}_{n}^2{\bf{I}}  \nonumber \\
& \ \ \ \ \ \ \ \ {\rm{Tr}}({\bf{X}}{\bf{X}}^{\rm{H}}) \le P.
\end{align}From the matrix theory perspective, Kronecker products usually do not affect the logic of deriving optimal solutions. It means for \textbf{Cases 3} and \textbf{4}, the procedures of deriving the optimal solutions are exactly the same.

\noindent {\textbf{(c) Dual-Hop AF MIMO Relaying Systems:}}

Recently, cooperative communications has received considerable research attention. Particularly, the joint design of the forwarding matrix at relay and the equalizer at destination has attracted a lot of attention \cite{Guan08,Tang07}. Robust transceiver designs are more complicated than their counterparts with perfect CSI and include the ones with perfect CSI as their special cases. The robust transceiver design maximizing mutual information between source and destination can be formulated as \cite{XingICSPCC}
\begin{align}
{\textbf{Case 5:}} \ \ & \min_{{\bf{X}}} \ \
{\rm{log}}|{\bf{A}}^{\rm{H}}({\bf{X}}^{\rm{H}}{\bf{H}}^{\rm{H}}{\bf{K}}_{\bf{X}}^{-1}{\bf{H}}{\bf{X}}
+{\bf{I}})^{-1}{\bf{A}}+{\bf{N}}|  \nonumber \\
& \ {\rm{s.t.}} \ \ \ {\bf{K}}_{\bf{X}}={\rm{Tr}}({\bf{X}}{\bf{X}}^{\rm{H}}{\boldsymbol \Psi}){\boldsymbol \Sigma}+{\sigma}_{n}^2{\bf{I}}  \nonumber \\
& \ \ \ \ \ \ \ \ {\rm{Tr}}({\bf{X}}{\bf{X}}^{\rm{H}}) \le P,
\end{align}where the variable ${\bf{X}}$ is the forwarding matrix at the relay and all the other matrices are constant. \textbf{Case 5} is similar to the optimization problem (24) in \cite{XingICSPCC}, except that the variable notation is changed from ${\bf{X}}$ to ${\bf{\tilde F}}$ and a basic constraint (i.e., ${\boldsymbol \Sigma}\propto {\bf{I}}$) is removed here.

Similar to point-to-point MIMO cases, when we focus on how accurately the desired signals can be recovered instead of how much information can be transmitted, MSE becomes to be a promising performance metric. For dual-hop AF MIMO relaying systems without source precoder design, the robust transceiver design minimizing MSE is equivalent to  \cite{Xing10}
\begin{align}
{\textbf{Case 6:}} \ \ & \min_{{\bf{X}}} \ \ {\rm{Tr}}[{\bf{A}}^{\rm{H}}({\bf{X}}^{\rm{H}}{\bf{H}}^{\rm{H}}{\bf{K}}_{\bf{X}}^{-1}{\bf{H}}{\bf{X}}
+{\bf{I}})^{-1}{\bf{A}}] \nonumber \\
& \ {\rm{s.t.}} \ \ \ {\bf{K}}_{\bf{X}}={\rm{Tr}}({\bf{X}}{\bf{X}}^{\rm{H}}{\boldsymbol \Psi}){\boldsymbol \Sigma}+{\sigma}_{n}^2{\bf{I}}  \nonumber \\
& \ \ \ \ \ \ \ \ {\rm{Tr}}({\bf{X}}{\bf{X}}^{\rm{H}}) \le P.
\end{align}Note that \textbf{Case 6} is corresponding to the optimization problem (33) in \cite{Xing10}. Although the formulations are significantly different, the nature is the same. The distinct difference is that when ${\boldsymbol \Psi}\not\propto {\bf{I}}$ and ${\boldsymbol \Sigma} \propto {\bf{I}}$, only a suboptimal solution is derived in \cite{Xing10}, while the exact optimal solution is given for this case in this paper.


\noindent {\textbf{(d) Additive Majorization Theory Based Optimization Problems:}}

To the best of our knowledge, for linear transceiver designs D. P. Palomar, et al. first discover that for various linear transceiver designs, the objective functions
can be written as some special functions of the diagonal elements of MSE matrix \cite{Palomar03}.
To unify various linear MIMO transceiver designs, the transceiver design can be formulated as an optimization problem with additive Schur-convex/concave functions of the diagonal elements of MSE matrix \cite{Palomar03}. Furthermore, this logic is also applicable to the robust cases. In the following we only focus on the robust cases that are more general \cite{Zhang08}. Specifically, with a given additive Schur-convex/concave objective function, the optimization problem of linear transceiver design is formulated as follows \cite{Zhang08}
\begin{align}
{\textbf{Case 7:}} \ \ & \min_{\bf{X}} \ \
{ f}_{\rm{A-Schur}}({\textbf{d}}[({\bf{X}}^{\rm{H}}{\bf{H}}^{\rm{H}}{\bf{K}}_{\bf{X}}^{-1}{\bf{H}}
{\bf{X}}+{\bf{I}})^{-1}])  \nonumber \\
& \ {\rm{s.t.}} \ \ \
{\bf{K}}_{\bf{X}}={\rm{Tr}}({\bf{X}}{\bf{X}}^{\rm{H}}{\boldsymbol \Psi}){\boldsymbol \Sigma}+{\sigma}_{n}^2{\bf{I}}  \nonumber \\
& \ \ \ \ \ \ \ \ {\rm{Tr}}({\bf{X}}{\bf{X}}^{\rm{H}}) \le P,
\end{align}where ${\textbf{d}}({\bf{Z}})$ represents the vector consisting of the diagonal elements of ${\bf{Z}}$, i.e., ${\textbf{d}}({\bf{Z}})=[[{\bf{Z}}]_{1,1},\cdots,
[{\bf{Z}}]_{N,N}]^{\rm{T}}$. Moreover, the objective function ${ f}_{\rm{A-Schur}}(\bullet)$ is increasing and additively Shur-concave/convex. Notice that the main difference between \cite{Zhang08} and our work is that in \cite{Zhang08} only the case of ${\boldsymbol \Psi} \propto {\bf{I}}$ is investigated, but in our work the case of ${\boldsymbol \Psi} \not\propto {\bf{I}}$ is also discussed.


\noindent {\textbf{(e) Multiplicative Majorization Theory Based Optimization Problems:}}

Similar to linear transceiver designs, for nonlinear transceiver designs with decision feedback equalizer (DFE) receiver and/or Tomlinson-Harashima precoding (THP) transmitter, the transceiver designs can be unified into an optimization problem with different objective functions that are increasing and multiplicative Schur-convex/concave \cite{Palomar2006}. Specifically, the optimization problem is given as
\begin{align}
{\textbf{Case 8:}} \ \ & \min \ \
{ f}_{\rm{M-Schur}}({\textbf{d}}^2[{\bf{L}}])  \nonumber \\
& \ {\rm{s.t.}} \ \ \ ({\bf{X}}^{\rm{H}}{\bf{H}}^{\rm{H}}{\bf{K}}_{\bf{X}}^{-1}{\bf{H}}
{\bf{X}}+{\bf{I}})^{-1}={\bf{L}}{\bf{L}}^{\rm{H}} \nonumber \\
&\ \ \ \ \ \ \ \ {\bf{K}}_{\bf{X}}={\rm{Tr}}({\bf{X}}{\bf{X}}^{\rm{H}}{\boldsymbol \Psi}){\boldsymbol \Sigma}+{\sigma}_{n}^2{\bf{I}}  \nonumber \\
& \ \ \ \ \ \ \ \ {\rm{Tr}}({\bf{X}}{\bf{X}}^{\rm{H}}) \le P,
\end{align}where ${\bf{L}}$ is a lower triangular matrix with nonnegative diagonal elements, which corresponds the Cholesky factorization of $({\bf{X}}^{\rm{H}}{\bf{H}}^{\rm{H}}{\bf{K}}_{\bf{X}}^{-1}{\bf{H}}
{\bf{X}}+{\bf{I}})^{-1}$. Moreover, ${\bf{d}}^2[{\bf{L}}]$ is a vector consisting of the square of the diagonal elements of ${\bf{L}}$, i.e., ${\bf{d}}^2[{\bf{L}}]=[[{\bf{L}}]_{1,1}^2 \dots [{\bf{L}}]_{N,N}^2]^{\rm{T}}$. Additionally, the objective function ${ f}_{\rm{M-Schur}}(\bullet)$ is increasing and multiplicative Schur-convex/concave. With perfect CSI, we simply have ${\boldsymbol \Psi}={\bf{0}}$ \cite{Shenouda2008,Amico2008}.

\noindent {\textbf{Remark:}} Training designs  and MIMO radar waveform optimizations are very similar and then due to space limitation we do not discuss the latter one in detail.

 All the above optimization problems from \textbf{Case 1} to  \textbf{8} can be unified as the following optimization problem
\begin{align}
\label{Opt_Original}
{\textbf{MVOP 1:}} \ \ & \min_{{\bf{X}}} \ \ \  {f}_{\rm{Matrix}}\left({\bf{K}}_{\bf{X}}^{-1/2}{\bf{H}}{\bf{X}}\right) \nonumber \\
& \ {\rm{s.t.}} \ \ \ \ {\bf{K}}_{\bf{X}}={\rm{Tr}}({\bf{X}}{\bf{X}}^{\rm{H}}{\boldsymbol \Psi}){\boldsymbol \Sigma}+{\sigma}_{n}^2{\bf{I}} \nonumber \\
 & \ \ \ \ \ \ \ \ \ {\rm{Tr}}({\bf{X}}{\bf{X}}^{\rm{H}}) \le P
\end{align} where the objective function is a matrix function $f_{\rm{Matrix}}(\bullet): {\mathbb{C}}^{M\times N}  \rightarrow {\mathbb{R}}$. Introducing an auxiliary unitary matrix ${\bf{Q}}$ and defining a new variable
\begin{align}
\label{Q}
{\bf{F}}={\bf{X}}{\bf{Q}}^{\rm{H}},
\end{align}
the objective function of  (\ref{Opt_Original}) can be rewritten as ${f}_{\rm{Matrix}}({\bf{K}}_{\bf{X}}^{-1/2}{\bf{H}}{\bf{X}})={f}_{\rm{Matrix}}({\bf{K}}_{\bf{F}}^{-1/2}
{\bf{H}}{\bf{F}}{\bf{Q}})$ where ${\bf{K}}_{\bf{F}}={\rm{Tr}}({\bf{F}}{\bf{F}}^{\rm{H}}{\boldsymbol \Psi}){\boldsymbol \Sigma}+{\sigma}_{n}^2{\bf{I}}$. Note that the unified optimization problem (\ref{Opt_Original}) has the following interesting and important property.

\noindent \textbf{Property 0:} For the optimal unitary matrix ${\bf{Q}}$, the matrix-variate objective function of (\ref{Opt_Original}) is minimized with respect to ${\bf{Q}}$ and it can be transformed to be a monotonic vector function, i.e.,
\begin{align}
 { f}_{\rm{Matrix}}\left({\bf{K}}_{\bf{F}}^{-1/2}
{\bf{H}}{\bf{F}}{\bf{Q}}\right) &\ge{ f}_{\rm{Matrix}}\left({\bf{K}}_{\bf{F}}^{-1/2}
{\bf{H}}{\bf{F}}{\bf{Q}}_{\rm{opt}}\right)={ f}_{\rm{Vector}}\left[{\boldsymbol \lambda}({\bf{F}}^{\rm{H}}{\bf{H}}^{\rm{H}}{\bf{K}}_{\bf{F}}^{-1}
{\bf{H}}{\bf{F}})\right],
\end{align}where ${ f}_{\rm{Vector}}(\bullet)$ is a decreasing vector function and ${\boldsymbol \lambda}({\bf{Z}})=
[\lambda_1({\bf{Z}}),\cdots,
\lambda_N({\bf{Z}})]^{\rm{T}}$ with $\lambda_n({\bf{Z}})$ denoting the $n^{\rm{th}}$ largest eigenvalue of ${\bf{Z}}$.

Notice that in \textbf{MVOP 1} there is no constraint on the newly introduced unitary matrix ${\bf{Q}}$. The derivation of the optimal ${\bf{Q}}$ should be investigated case by case. Due to the case-dependency of the derivation, the discussion on ${\bf{Q}}_{\rm{opt}}$ will be deferred to Section IV. Based on the optimal ${\bf{Q}}$, the optimization problem (\ref{Opt_Original}) becomes the following optimization problem
\begin{align}
\label{Eigen_Opt}
{\textbf{MVOP 2:}} \ \  & \min_{{\bf{F}}} \ \ \ { f}_{\rm{Vector}}\left[{\boldsymbol \lambda}({\bf{F}}^{\rm{H}}{\bf{H}}^{\rm{H}}{\bf{K}}_{\bf{F}}^{-1}
{\bf{H}}{\bf{F}})\right] \nonumber \\
& \ {\rm{s.t.}} \ \ \ \ {\bf{K}}_{\bf{F}}={\rm{Tr}}({\bf{F}}{\bf{F}}^{\rm{H}}{\boldsymbol \Psi}){\boldsymbol \Sigma}+{\sigma}_{n}^2{\bf{I}} \nonumber \\
 & \ \ \ \ \ \ \ \ \ {\rm{Tr}}({\bf{F}}{\bf{F}}^{\rm{H}}) \le P,
\end{align}which can be regarded as an MMOP and will be the focus of the following section.

\section{Fundamentals of Matrix-Monotonic Optimization}
\label{Sect: Fundamentals}

\subsection{Matrix-Monotonic Optimization Problem}

\noindent \textbf{Definition 1:} A matrix-monotone function is defined to be a function  ${f}(\bullet)$ which maps a matrix variable from a subset of Hermitian matrices
 to a real number. If ${f}(\bullet)$ is a matrix-monotone decreasing function on positive semi-definite matrices, it satisfies
\begin{align}
{\boldsymbol{N}} \succeq {\boldsymbol M}\succeq {\bf{0}} \rightarrow {f}({\boldsymbol{N}}) \le {f}({\boldsymbol{M}}).
\end{align}On the other hand, when ${f}(\bullet)$ is matrix-monotone increasing, it means $-{f}(\bullet)$ is a matrix-monotone decreasing function \cite{Jorswieck07}.

Based on the above definition and together with the properties of positive semi-definite matrices \cite{Horn85}, the function ${ f}_{\rm{Vector}}\left[{\boldsymbol \lambda}(\bullet)\right]$ in (\ref{Eigen_Opt}) has the following two important properties that can be exploited further to derive the optimal solutions.

\noindent \textbf{Property 1:} The function ${ f}_{\rm{Vector}}\left[{\boldsymbol \lambda}({\boldsymbol M})\right]$ is a matrix-monotone decreasing function with respect to the matrix variable ${\boldsymbol M}$.

\noindent \textbf{Property 2:}  The function ${ f}_{\rm{Vector}}\left[{\boldsymbol \lambda}({\boldsymbol M})\right]$ is independent of the unitary matrix of the eigenvalue decomposition (EVD) of ${\boldsymbol M}$.


Based on \textbf{Property 1} and \textbf{2}, for all possible objective functions $f_{\rm{Vector}}(\bullet)$ the optimization problem (\ref{Eigen_Opt}) is in nature equivalent to  the following multi-objective optimization problem which is named as MMOP in our work
\begin{align}
\label{matrix_m_opt}
{\textbf{MMOP 1:}} \ \ & \max_{{\bf{F}}} \ \   {\bf{F}}^{\rm{H}}{\bf{H}}^{\rm{H}}
{\bf{K}}_{\bf{F}}^{-1}
{\bf{H}}{\bf{F}} \nonumber \\
& \ {\rm{s.t.}} \ \ \ \ {\bf{K}}_{\bf{F}}={\rm{Tr}}({\bf{F}}{\bf{F}}^{\rm{H}}{\boldsymbol \Psi}){\boldsymbol \Sigma}+{\sigma}_{n}^2{\bf{I}} \nonumber \\
 & \ \ \ \ \ \ \ \ \ {\rm{Tr}}({\bf{F}}{\bf{F}}^{\rm{H}}) \le P.
\end{align}
The most distinct characteristic of the optimization problem (\ref{matrix_m_opt}) is that the objective function is a positive semi-definite \textbf{matrix} instead of a real scalar. The maximization operation is defined based on monotonicity in positive semi-definite matrices.
This kind of optimization problems belongs to the category of multi-objective optimization problems \cite[P177]{Boyd04}, \cite{Ehrgott2005}. It should also be pointed out that the equivalence between the two optimization problems (\ref{Eigen_Opt}) and (\ref{matrix_m_opt}) is built on the whole set of all possible objective functions in (\ref{Eigen_Opt}). For a given $f_{\rm{Vector}}(\bullet)$ with a clear and specific formulation, the set of the solutions of (\ref{Eigen_Opt}) is a subset of the one of (\ref{matrix_m_opt}). In a unified version, the set of the solutions of all possible $f_{\rm{Vector}}(\bullet)$ will be equivalent to that of (\ref{matrix_m_opt}).

The objective function of maximizing $ {\bf{F}}^{\rm{H}}{\bf{H}}^{\rm{H}}
{\bf{K}}_{\bf{F}}^{-1}
{\bf{H}}{\bf{F}} $ implies the following  two important properties:

\noindent \textbf{Property 3:} In nature the objective function of (\ref{matrix_m_opt}) is to maximize the vector of the eigenvalues of $ {\bf{F}}^{\rm{H}}{\bf{H}}^{\rm{H}}
{\bf{K}}_{\bf{F}}^{-1}
{\bf{H}}{\bf{F}} $, i.e., $ {\boldsymbol \lambda}({\bf{F}}^{\rm{H}}{\bf{H}}^{\rm{H}}
{\bf{K}}_{\bf{F}}^{-1}
{\bf{H}}{\bf{F}})$ \cite[P. 177]{Boyd04}.

\noindent \textbf{Property 4:} The unitary matrix of the EVD of $ {\bf{F}}^{\rm{H}}{\bf{H}}^{\rm{H}}
{\bf{K}}_{\bf{F}}^{-1}
{\bf{H}}{\bf{F}} $ does not affect the objective. In other words, if ${\bf{F}}_{\rm{opt}}$ is one of Pareto optimal points and ${\bf{U}}$ is an unitary matrix, ${\bf{F}}_{\rm{opt}}{\bf{U}}$ is still a Pareto optimal solution \cite[P. 177]{Boyd04}.

Based on these two properties, we can have two useful lemmas that can help us simplify the considered optimization problem.

\noindent \textbf{{\textsl{Lemma 1:}}} As ${f}_{\rm{Monot}}(\bullet)$ is a matrix-monotone decreasing function on positive semi-definite matrices, the optimal solution of the optimization problem (\ref{matrix_m_opt}) always occurs on the boundary. In other words, a necessary condition for the optimal solutions is
\begin{align}
{\rm{Tr}}({\bf{F}}{\bf{F}}^{\rm{H}})=P.
\end{align}

\noindent \textsl{Proof:} The proof can be found in Appendix C in\cite{XingTSP2013}. $\blacksquare$


\noindent \textbf{{\textsl{Lemma 2:}}} Defining an auxiliary variable
\begin{align}
\label{eta_f}
\eta_f \triangleq {\rm{Tr}}({\bf{F}}{\bf{F}}^{\rm{H}}{\boldsymbol \Psi})\alpha+\sigma_{n}^2 \ \ \text{with} \ \ \alpha=\lambda_{\min}({\boldsymbol \Sigma}),
\end{align}$ {\rm{Tr}}({\bf{F}}{\bf{F}}^{\rm{H}})=P$ is equivalent to
$
 {\rm{Tr}}[{\bf{F}}
{\bf{F}}^{\rm{H}}(\alpha P{\boldsymbol \Psi}+\sigma_n^2{\bf{I}})]/\eta_f=P$.

\noindent \textsl{Proof:} The proof can be found in Appendix C in \cite{XingTSP2013}. $\blacksquare$

Based on \textbf{Lemma 1} and \textbf{2}, the optimization problem (\ref{matrix_m_opt}) is equivalent to the following optimization problem
\begin{align}
\label{opt:app_3_1}
\textbf{MMOP 2:} \ \ \ \ & \max_{{\bf{F}}} \ \ {\bf{F}}^{\rm{H}}{\bf{H}}^{\rm{H}}{\bf{K}}_{\bf{F}}^{-1}
{\bf{H}}{\bf{F}} \nonumber \\
& \ {\rm{s.t.}} \ \ \ \ {\bf{K}}_{\bf{F}}={\rm{Tr}}({\bf{F}}{\bf{F}}^{\rm{H}}{\boldsymbol \Psi}){\boldsymbol \Sigma}+{\sigma}_{n}^2{\bf{I}}\nonumber \\
& \ \ \ \ \ \ \ \ \ {\rm{Tr}}[{\bf{F}}{\bf{F}}^{\rm{H}}(\alpha P{\boldsymbol
\Psi}+\sigma_{n}^2{\bf{I}})]/\eta_{f}=P.
\end{align}In order to further simplify the mathematical formulation, the below new variable is defined
\begin{align}
\label{definitions_F_Pi}
{\bf{\tilde F}}&\triangleq {1}/{\sqrt{\eta_f}}(\alpha P{\boldsymbol
\Psi}+\sigma_{n}^2{\bf{I}})^{1/2}{\bf{F}},
\end{align} and then the optimization problem (\ref{opt:app_3_1}) is reformulated as
\begin{align}
\label{Matrix Optimization Problem}
 & \max_{{\bf{\tilde F}}} \ \  {\bf{\tilde F}}^{\rm{H}}(\alpha P{\boldsymbol
\Psi}+\sigma_{n}^2{\bf{I}})^{-1/2}{\bf{H}}^{\rm{H}}({\bf{K}}_{\bf{F}}/\eta_{f})^{-1}
{\bf{H}}(\alpha P{\boldsymbol
\Psi}+\sigma_{n}^2{\bf{I}})^{-1/2}{\bf{\tilde F}} \nonumber \\
& \ {\rm{s.t.}} \ \ \ \  {\rm{Tr}}({\bf{\tilde F}}{\bf{\tilde F}}^{\rm{H}})= P.
\end{align}
For the case with ${\boldsymbol{ \Psi}}{\not\propto} {\bf{I}}$ and ${\boldsymbol \Sigma}{\not\propto} {\bf{I}}$, ${\bf{K}}_{\bf{F}}/\eta_{f}$ is a complicated function of ${\bf{F}}$ and thus the structure of the optimal solution is difficult to derive. To the best of our knowledge, even for a simple MSE minimization problem, the closed-form optimal solution for this case is still largely open \cite{Ding09,Zhang08}.  In the following, we mainly focus on the cases with ${\boldsymbol{ \Psi}}{\propto} {\bf{I}}$ or ${\boldsymbol \Sigma}{\propto} {\bf{I}}$, which have clear and important physical meanings \cite{Ding09,Zhang08,Xing10}. For example, it has been proved in \cite{Ding09} that for a practical channel estimation algorithm, if the training power is large enough, ${\boldsymbol{ \Psi}}{\propto} {\bf{I}}$ asymptotically holds. In practice, this training power assumption is very reasonable.

Defining the following two constant matrices
\begin{align}
\label{upper_bound}
{\bf{K}}_{{\boldsymbol \Psi}} & \triangleq {\frac{P\lambda_{\max}({\boldsymbol \Psi})}{{P\lambda_{\max}({\boldsymbol \Psi})\alpha+\sigma_{n}^2}}{\boldsymbol \Sigma}+\frac{\sigma_{n}^2}{{P\lambda_{\max}({\boldsymbol \Psi})}\alpha+\sigma_{n}^2}{\bf{I}}} \nonumber \\
{\boldsymbol \Pi} & \triangleq {\bf{K}}_{\boldsymbol{\Psi}}^{-1/2}
{\bf{H}}(\alpha P{\boldsymbol
\Psi}+\sigma_{n}^2{\bf{I}})^{-1/2},
\end{align} when ${\boldsymbol{ \Psi}}{\propto} {\bf{I}}$ or ${\boldsymbol \Sigma}{\propto} {\bf{I}}$, the optimization problem (\ref{Matrix Optimization Problem}) is equivalent to the following optimization problem
\begin{align}
\label{MMOP3}
\textbf{MMOP 3:} \ \ \ \ & \max_{{\bf{\tilde F}}} \ \  {\bf{\tilde F}}^{\rm{H}}{\boldsymbol \Pi}^{\rm{H}}{\boldsymbol \Pi}{\bf{\tilde F}} \nonumber \\
& \ {\rm{s.t.}} \ \ \ \ {\rm{Tr}}({\bf{\tilde F}}{\bf{\tilde F}}^{\rm{H}})\le P.
\end{align}

Note that $ {\bf{\tilde F}}^{\rm{H}}{\boldsymbol \Pi}^{\rm{H}}{\boldsymbol \Pi}{\bf{\tilde F}}$ is convex over the set of positive semi-definite matrices with respect to  ${\bf{\tilde F}}$ \cite{Boyd04}. Relying on multi-objective optimization theory, the solution of ${\bf{\tilde F}}$ of the following optimization problem is a Pareto optimal solution of the optimization problem (\ref{Matrix Optimization Problem}) \cite{Boyd04}
\begin{align}
\label{optmal_prop}
& \max_{\rho,{\bf{\tilde F}} } \ \ \ \ \ \ \ \  \rho \nonumber \\
& \  {\rm{s.t.}} \ \ \ \ {\rm{Tr}}({\bf{\tilde F}}{\bf{\tilde F}}^{\rm{H}})\le  P \nonumber \\
& \ \ \ \ \ \ \ \ \ {\bf{\tilde F}}^{\rm{H}}{\boldsymbol{\Pi}}^{\rm{H}}{\boldsymbol{\Pi}}{\bf{\tilde F}}=\rho {\bf{\tilde F}}_{\rm{in}}^{\rm{H}}{\boldsymbol{\Pi}}^{\rm{H}}{\boldsymbol{\Pi}}{\bf{\tilde F}}_{\rm{in}}
\end{align}where ${\bf{\tilde F}}_{\rm{in}}$ is an inner point in the ball ${\rm{Tr}}({\bf{\tilde F}}{\bf{\tilde F}}^{\rm{H}})\le P$, i.e., ${\rm{Tr}}({\bf{\tilde F}}_{\rm{in}}{\bf{\tilde F}}_{\rm{in}}^{\rm{H}})< P$. The symbol $\rho$ is a real positive variable. It should be noticed that the whole Pareto optimal set of ${\bf{\tilde F}}$ of  problem (\ref{Matrix Optimization Problem})  can be achieved via changing ${\bf{\tilde F}}_{\rm{in}}$. In Appendix~\ref{App:Conclusion}, the structure of the optimal solution of the optimization problem (\ref{optmal_prop}) is derived. It is discovered that the structure is independent of ${\bf{\tilde F}}_{\rm{in}}$. It means that all points in the Pareto optimal set have the same structure which is exactly the structure of the optimal solution of  (\ref{Matrix Optimization Problem}).

Defining a unitary matrix ${\bf{V}}_{{\boldsymbol{\Pi}}}$ and a rectangular diagonal matrix ${\boldsymbol \Lambda}_{{\boldsymbol{\Pi}}}$ based on the following singular value decomposition (SVD)
\begin{align}
&{\boldsymbol{\Pi}}={\bf{U}}_{\boldsymbol{\Pi}}{\boldsymbol \Lambda}_{\boldsymbol{\Pi}}
{\bf{V}}_{\boldsymbol{\Pi}}^{\rm{H}} \ \ {\text{with}} \ \ {\boldsymbol \Lambda}_{\boldsymbol{\Pi}} \searrow
\end{align} we can prove the following two important lemmas.

$ \ $

\noindent \textbf{{\textsl{Lemma 3:}}} The optimal solution of the optimization problem (\ref{optmal_prop}) has the following structure
\begin{align}
&{\bf{\tilde F}}_{\rm{opt}}={\bf{V}}_{\boldsymbol{\Pi}}{\boldsymbol \Lambda}_{\bf{F}}{\bf{U}}_{\rm{Arb}}^{\rm{H}}
\end{align}where the matrix ${\boldsymbol \Lambda}_{\bf{F}}$ is a rectangular diagonal matrix and the matrix ${\bf{U}}_{\rm{Abr}}$ is an arbitrary unitary matrix with a proper dimension.

\noindent \textbf{{\textsl{Lemma 4:}}} The diagonal matrix ${\boldsymbol \Lambda}_{\bf{F}}$ must satisfy the following property
\begin{align}
{\boldsymbol \Lambda}_{\bf{F}}^{\rm{H}}{\boldsymbol \Lambda}_{\boldsymbol{\Pi}}^{\rm{H}}{\boldsymbol \Lambda}_{\boldsymbol{\Pi}}{\boldsymbol \Lambda}_{\bf{F}} \searrow.
\end{align}

\noindent \textbf{\textsl{Proof:}} See Appendix~\ref{App:Conclusion}. $\blacksquare$

$ \ $

It is clear that the results shown in \textbf{Lemmas 3} and \textbf{4} are  independent of ${\bf{\tilde F}}_{\rm{in}}$\footnote{It should be highlighted that only the structure is independent of ${\bf{\tilde F}}_{\rm{in}}$.}. In other words, all Pareto optimal points must satisfy \textbf{Lemmas 3} and \textbf{4}.
Then using the definition of ${\bf{\tilde F}}$, it follows that
\begin{align}
\label{Apped_X_OPT}
{\bf{F}}_{\rm{opt}}=\sqrt{\eta_f}(\alpha P{\boldsymbol
\Psi}+\sigma_{n}^2{\bf{I}})^{-1/2}{\bf{V}}_{\boldsymbol{\Pi}}{\boldsymbol \Lambda}_{\bf{F}}{\bf{U}}_{\rm{Arb}}^{\rm{H}}.
\end{align}Substituting (\ref{Apped_X_OPT}) into the definition of $\eta_f$ in (\ref{eta_f}), we can achieve a simple linear function of ${\eta}_f$, and $\eta_f$ can be easily solved as
\begin{align}
\eta_f&={\sigma_{n}^2}/\{1-\alpha{\rm{Tr}}
[{\bf{V}}_{{\boldsymbol{\Pi}}}^{\rm{H}}(\alpha P{\boldsymbol
\Psi}+\sigma_{n}^2{\bf{I}})^{-{\rm{H}}/2}{\boldsymbol
\Psi}(\alpha P{\boldsymbol
\Psi}+\sigma_{n}^2{\bf{I}})^{-1/2}
{\bf{V}}_{{\boldsymbol{\Pi}}}{\boldsymbol{\Lambda}}_{{\bf{F}}}{\boldsymbol{\Lambda}}_{{\bf{F}}}^{\rm{H}}]\} \nonumber \\
&=P/{\rm{Tr}}[(\alpha P{\boldsymbol
\Psi}+\sigma_{n}^2{\bf{I}})^{-1/2}{\bf{V}}_{\boldsymbol{\Pi}}{\boldsymbol{\Lambda}}_{{\bf{F}}}
{\boldsymbol{\Lambda}}_{{\bf{F}}}^{\rm{H}}{\bf{V}}_{\boldsymbol{\Pi}}^{\rm{H}}
(\alpha P{\boldsymbol
\Psi}+\sigma_{n}^2{\bf{I}})^{-{\rm{H}}/2}].
\end{align} The above results can be summarized as the following theorem.

$ \ $

\noindent \textbf{{\textsl{Theorem 1:}}} When ${\boldsymbol{ \Psi}}{\propto} {\bf{I}}$ or ${\boldsymbol \Sigma}{\propto} {\bf{I}}$, the optimal solution of the optimization problem (\ref{matrix_m_opt}) has the following structure
\begin{align}
\label{conclusion_2}
&{\bf{F}}_{\rm{opt}}=\sqrt{\eta_f}(\alpha P{\boldsymbol
\Psi}+\sigma_{n}^2{\bf{I}})^{-1/2}{\bf{V}}_{\boldsymbol{\Pi}}{\boldsymbol \Lambda}_{\bf{F}}{\bf{U}}_{\rm{Arb}}^{\rm{H}}
\end{align}where the scalar $\eta_f$ equals
\begin{align}
\eta_f=P/{\rm{Tr}}[(\alpha P{\boldsymbol
\Psi}+\sigma_{n}^2{\bf{I}})^{-1}{\bf{V}}_{\boldsymbol{\Pi}}{\boldsymbol{\Lambda}}_{{\bf{F}}}
{\boldsymbol{\Lambda}}_{{\bf{F}}}^{\rm{H}}{\bf{V}}_{\boldsymbol{\Pi}}^{\rm{H}}
].
\end{align}

\noindent \textbf{Remark:} When ${\boldsymbol{ \Psi}}{\propto} {\bf{I}}$ or ${\boldsymbol \Sigma}{\propto} {\bf{I}}$, (\ref{conclusion_2}) is exactly the optimal solution. However, if ${\boldsymbol{ \Psi}} {\not\propto} {\bf{I}}$ and ${\boldsymbol \Sigma} {\not\propto} {\bf{I}}$, (\ref{conclusion_2}) becomes a suboptimal solution to (\ref{Matrix Optimization Problem}). It is worth noting that in this case (\ref{conclusion_2}) is still meaningful. Based on the definition in (\ref{upper_bound}), it can be concluded that ${\bf{K}}_{{\bf{F}}}/\eta_{f} \preceq {\bf{K}}_{\boldsymbol \Psi}$. Then it can be proved that $ {\bf{\tilde F}}^{\rm{H}}{\boldsymbol \Pi}^{\rm{H}}{\boldsymbol \Pi}{\bf{\tilde F}}$ is a lower bound of the objective function of (\ref{Matrix Optimization Problem}). As (\ref{Matrix Optimization Problem}) aims at maximizing the objective function, a lower bound is meaningful and then (\ref{conclusion_2}) is still a meaningful suboptimal solution.

\subsection{Optimal Solution of ${\boldsymbol{\Lambda}}_{\bf{F}}$}

Based on the optimal structure given by \textbf{Theorem 1}, the only remaining variable to be optimized is the rectangular diagonal matrix ${\boldsymbol \Lambda}_{\bf{F}}$. Generally speaking, different optimization problems usually have different formulations of ${\boldsymbol \Lambda}_{\bf{F}}$, but their derivation procedures are almost the same, which mainly rely on KKT conditions. In particular, in most cases the optimal solutions of ${\boldsymbol \Lambda}_{\bf{F}}$ are different variants of classic water-filling solutions. In this section, we derive the optimal solutions of ${\boldsymbol \Lambda}_{\bf{F}}$ from the perspective of multi-objective optimization theory.

Substituting the optimal structure into the original optimization problem (\ref{MMOP3}) and defining $[{\boldsymbol {\Lambda}}_{{\boldsymbol \Pi}}]_{n,n}=\lambda_{{\boldsymbol \Pi},n}$ and $[{\boldsymbol {\Lambda}}_{{\bf{F}}}]_{n,n}=f_{n}$, we directly have the following vector optimization problem
\begin{align}
\label{MOLP}
\textbf{MOLP:} \ \ \ \ & \max_{\{f_n^2\}} \ \    [\lambda_{{\boldsymbol \Pi},1}^2 f_1^2,\cdots, \lambda_{{\boldsymbol \Pi},N}^2 f_N^2]^{\rm{T}} \nonumber \\
& \ {\rm{s.t.}} \ \ \  \lambda_{{\boldsymbol \Pi},1}^2 f_1^2 \ge  \lambda_{{\boldsymbol \Pi},2}^2 f_2^2 \cdots \ge \lambda_{{\boldsymbol \Pi},N}^2 f_N^2 \nonumber \\
& \ \ \ \ \ \ \  \sum_{n=1}^N f_n^2 \le P.
\end{align}Taking $f_n^2$'s as the variables, it is obvious that the optimization problem (\ref{MOLP}) is multi-objective linear programming which can be efficiently solved by numerical methods \cite{Boyd04}. However, closed-form solutions are generally preferred for complexity reduction. In order to achieve closed-form solutions, certain transformations are necessary. The set of  Pareto optimal solutions of the optimization problem (\ref{MOLP}) is exactly the Pareto optimal point set of the following optimization problem without the order constraint \cite{Boyd04}
\begin{align}
\label{Vector_Optimization_aaa}
& \min_{\{f_n^2\}} \ \    \left[\frac{1}{\lambda_{{\boldsymbol \Pi},1}^2 f_1^2+1},\cdots, \frac{1}{\lambda_{{\boldsymbol \Pi},N}^2 f_N^2+1}\right]^{\rm{T}} \nonumber \\
& \ {\rm{s.t.}} \ \ \ \sum_{n=1}^N f_n^2 \le P,
\end{align} in which the linear functions of $\{f_n^2\}$ in \textbf{MOLP} have been replaced by convex functions. The order constraint is removed because it can be simply realized by choosing proper weighting factors. As ${1}/({\lambda_{{\boldsymbol \Pi},n}^2 f_n^2+1})$
is a convex function with respect to $f_n^2$, the Pareto solutions can be efficiently achieved by solving its scalarization problem \cite{Boyd04}. Notice that if other convex functions are used, e.g., $-{\rm{log}}({\lambda_{{\boldsymbol \Pi},n}^2 f_n^2+1})$, similar results can still be achieved. Introducing a series of positive weight factors $w_n$'s, the Pareto optimal solutions  of (\ref{Vector_Optimization_aaa}) can be attained by solving the following single objective optimization problem
\begin{align}
& \min_{\{f_n^2\}} \ \    \sum_n \frac{w_n}{\lambda_{{\boldsymbol \Pi},n}^2 f_n^2+1} \nonumber \\
& \ {\rm{s.t.}} \ \ \ \sum_{k=1}^K f_k^2 \le P,
\end{align}whose optimal solution is water-filling solution, i.e.,
\begin{align}
\label{water_filling}
f_n^2=\left(\sqrt{\frac{w_n }{\lambda_{{\boldsymbol \Pi},n}^2 \mu}}-\frac{1}{\lambda_{{\boldsymbol \Pi},n}^2}\right)^+.
\end{align}In order to guarantee that $\lambda_{{\boldsymbol \Pi},1}^2 f_1^2 \ge  \lambda_{{\boldsymbol \Pi},2}^2 f_2^2 \cdots \ge \lambda_{{\boldsymbol \Pi},N}^2 f_N^2$, the weighting factors must satisfy
\begin{align}
\label{weighting_order}
\sqrt{w_1{\lambda_{{\boldsymbol \Pi},1}^2 }}\ge  \sqrt{w_2{\lambda_{{\boldsymbol \Pi},2}^2 }} \cdots \ge \sqrt{w_N{\lambda_{{\boldsymbol \Pi},N}^2 }}.
\end{align}

It has been shown in \textbf{Theorem 1} that the formulated MMOP has some elegant properties for finding the optimal solution. In MIMO system design, however another challenging problem is to transform the MVOP in (\ref{Opt_Original}) into the MMOP in (\ref{Eigen_Opt}), i.e., to find the optimal unitary matrix ${\bf{Q}}$ in (\ref{Q}). This issue will be investigated in depth in the next section.

\noindent \textbf{Remark:} An important dual optimization problem of (\ref{optmal_prop}) is
\begin{align}
& \min_{{\bf{\tilde F}}} \ \ \ \ \ {\rm{Tr}}({\bf{\tilde F}}{\bf{\tilde F}}^{\rm{H}}) \nonumber \\
& \  {\rm{s.t.}} \ \ \ \ \ \ {\bf{\tilde F}}^{\rm{H}}{\boldsymbol{{\Pi}}}^{\rm{H}}{\boldsymbol{{\Pi}}}{\bf{\tilde F}}= {\bf{\tilde F}}_{\rm{Pareto}}^{\rm{H}}{\boldsymbol{{\Pi}}}^{\rm{H}}{\boldsymbol{{\Pi}}}{\bf{\tilde F}}_{\rm{Pareto}}. \nonumber
\end{align}This optimization problem usually corresponds to QoS-based designs. e.g., minimizing the transmit power subject to a certain performance requirement. It is worth noting that the dual optimization problem has the same solution as that given in \textbf{Theorem 1}.




\section{Unitary-Matrix Based Transformations to MMOPs}
\label{Sect: Unitary_Transform}

Generally speaking, the computation of the optimal unitary matrix ${\bf{Q}}$ relies on matrix inequality theory and depends on the objective functions \cite{Marshall79}. In the following, two kinds of powerful methods, i.e., basic matrix inequality based method and Majorization theory based method, are introduced to derive the optimal solutions of ${\bf{Q}}$ under different objective functions.

\subsection{Basic Matrix Inequality Based Method}

\noindent \textbf{{\underline{Basic Matrix Inequalities:}}}

 Given two $N \times N$ positive semi-definite matrices ${\boldsymbol {A}}$ and ${\boldsymbol {B}}$, we can have the following EVDs
\begin{align}
&{\boldsymbol {A}}={\bf{U}}_{{\boldsymbol {A}}}{\boldsymbol \Lambda}_{{\boldsymbol {A}}}{\bf{U}}_{{\boldsymbol {A}}}^{\rm{H}}={\bf{\bar U}}_{{\boldsymbol {A}}}{\boldsymbol {\bar \Lambda}}_{{\boldsymbol {A}}}{\bf{\bar U}}_{{\boldsymbol {A}}}^{\rm{H}} \ \ \text{with} \ \ {\boldsymbol \Lambda}_{{\boldsymbol {A}}} \searrow \ \text{and} \ \ {\boldsymbol {\bar \Lambda}}_{{\boldsymbol {A}}}\nearrow \\
&{\boldsymbol {B}}={\bf{U}}_{{\boldsymbol {B}}}{\boldsymbol \Lambda}_{{\boldsymbol {B}}}{\bf{U}}_{{\boldsymbol {B}}}^{\rm{H}}={\bf{\bar U}}_{{\boldsymbol {B}}}{\boldsymbol {\bar \Lambda}}_{{\boldsymbol {B}}}{\bf{\bar U}}_{{\boldsymbol {B}}}^{\rm{H}} \ \ \text{with} \ \ {\boldsymbol \Lambda}_{{\boldsymbol {B}}} \searrow \ \text{and} \ \ {\boldsymbol {\bar \Lambda}}_{{\boldsymbol {B}}}\nearrow,
\end{align}based on which four basic matrix inequalities are given first in the following. They are the theoretical basis for the derivation of the optimal solution of ${\bf{Q}}$.

\noindent \textbf{\textsl{Inequality 1:}} The trace of a product of two positive semi-definite matrices satisfies  \cite{Marshall79}
\begin{align}
& {\sum}_{i=1}^N\lambda_i({\boldsymbol A})\lambda_{N-i+1}({\boldsymbol B})  \le {\rm{Tr}}({\bf{Q}}^{\rm{H}}{\boldsymbol A}{\bf{Q}}{\boldsymbol B}) \le {\sum}_{i=1}^N\lambda_i({\boldsymbol A})\lambda_i({\boldsymbol B}) \label{inequ_1}
\end{align}where the left equality holds when the unitary matrix ${\bf{Q}}$ satisfies
${\bf{Q}}={\bf{U}}_{{\boldsymbol {A}}}{\bf{\bar U}}_{{\boldsymbol {B}}}^{\rm{H}}$ and the right equality holds when
${\bf{Q}}={\bf{U}}_{{\boldsymbol {A}}}{\bf{ U}}_{{\boldsymbol {B}}}^{\rm{H}}$.

\noindent \textbf{\textsl{Inequality 2:}} The determinant of the sum of positive semi-definite matrices satisfies \cite{Marshall79}
\begin{align}
& {\prod}_{i=1}^N(\lambda_i({\boldsymbol A})+\lambda_{i}({\boldsymbol B}))  \le |{\bf{Q}}^{\rm{H}}{\boldsymbol A}{\bf{Q}}+{\boldsymbol B}| \le {\prod}_{i=1}^N(\lambda_i({\boldsymbol A})+\lambda_{N-i+1}({\boldsymbol B})), \label{inequ_2} \end{align}where the left equality holds when
${\bf{Q}}={\bf{U}}_{{\boldsymbol {A}}}{\bf{ U}}_{{\boldsymbol {B}}}^{\rm{H}}$ and the right equality holds when
${\bf{Q}}={\bf{U}}_{{\boldsymbol {A}}}{\bf{\bar U}}_{{\boldsymbol {B}}}^{\rm{H}}$.

\noindent \textbf{\textsl{Inequality 3:}} For positive semi-definite matrices ${\boldsymbol {A}}$ and ${\boldsymbol {B}}$, we have
\begin{align}
& {\prod}_{i=1}^N(\lambda_i({\boldsymbol {A}}) \lambda_{N-i+1}({\boldsymbol {B}})+1)\le|{\bf{Q}}^{\rm{H}}{\boldsymbol {A}}{\bf{Q}}{\boldsymbol B}+{\bf{I}}|\le {\prod}_{i=1}^N(\lambda_i({\boldsymbol {A}}) \lambda_{i}({\boldsymbol {B}})+1)\label{inequ_3}
\end{align}where the left equality holds when
${\bf{Q}}={\bf{U}}_{{\boldsymbol {A}}}{\bf{\bar U}}_{{\boldsymbol {B}}}^{\rm{H}}$. On the other hand, the right equality holds when
${\bf{Q}}={\bf{U}}_{{\boldsymbol {A}}}{\bf{ U}}_{{\boldsymbol {B}}}^{\rm{H}}$.

\noindent \textbf{\textsl{Inequality 4:}} Given two positive semi-definite matrices ${\boldsymbol {A}}$ and ${\boldsymbol {B}}$, the following relationship holds
\begin{align}
& {\sum}_{i=1}^N(\lambda_i({\boldsymbol {A}})+ \lambda_{N-i+1}({\boldsymbol {B}}))^{-1}\le {\rm{Tr}}[({\bf{Q}}^{\rm{H}}{\boldsymbol {A}}{\bf{Q}}+{\boldsymbol B})^{-1}]\le {\sum}_{i=1}^N(\lambda_i({\boldsymbol {A}})+ \lambda_i({\boldsymbol {B}}))^{-1}\label{inequ_4}
\end{align}where the left equality holds when
${\bf{Q}}={\bf{U}}_{{\boldsymbol {A}}}{\bf{\bar U}}_{{\boldsymbol {B}}}^{\rm{H}}$ and the right equality holds when
${\bf{Q}}={\bf{U}}_{{\boldsymbol {A}}}{\bf{ U}}_{{\boldsymbol {B}}}^{\rm{H}}$.

\noindent \textbf{\textsl{Proof:}} The proofs of Inequalities 1,  2, and  3 can be found in \cite{Marshall79}, while the proof of Inequality 4 is shown in Appendix~\ref{App:Proof_Inequ}. Notice that following the logic of the proof for Inequality 4, the first three inequalities can also be proved.  $\blacksquare$

\noindent {\textbf{{\underline{Conclusions for Cases 1 to 6:}}}}

Given a complex matrix ${\bf{A}}$ and a positive definite matrix ${\bf{N}}$, based on the following SVDs and EVDs
\begin{align}
\label{definitions}
& {\bf{A}}={\bf{U}}_{\bf{A}}{\boldsymbol \Lambda}_{\bf{A}}{\bf{V}}_{\bf{A}}^{\rm{H}} \ \ {\text{with}} \ \ {\boldsymbol \Lambda}_{\bf{A}} \searrow \nonumber \\
&  {\bf{A}}{\bf{N}}^{-1}{\bf{A}}^{\rm{H}}={\bf{U}}_{\bf{ANA}}{\boldsymbol \Lambda}_{\bf{ANA}}
{\bf{U}}_{\bf{ANA}}^{\rm{H}}\ \ {\text{with}} \ \ {\boldsymbol \Lambda}_{\bf{ANA}} \searrow \nonumber \\
& {\bf{N}}={\bf{\bar U}}_{\bf{N}}{\boldsymbol {\bar \Lambda}}_{\bf{N}}
{\bf{\bar U}}_{\bf{N}}^{\rm{H}}\ \ {\text{with}} \ \ {\boldsymbol {\bar \Lambda}}_{\bf{N}} \nearrow, \nonumber \\
&{\bf{F}}^{\rm{H}}{\bf{H}}^{\rm{H}}{\bf{K}}_{\bf{F}}^{-1}
{\bf{H}}{\bf{F}}={\bf{U}}_{\bf{FHF}}{\boldsymbol \Lambda}_{\bf{FHF}}
{\bf{U}}_{\bf{FHF}}^{\rm{H}}\ \ {\text{with}} \ \ {\boldsymbol \Lambda}_{\bf{FHF}} \searrow
\end{align} and together with the above four inequalities, the following results can be achieved.

\noindent {\textbf{\textsl{Conclusion 1:}}} Based on (\ref{definitions}) and Inequality 2, ${\bf{Q}}_{\rm{opt}}$ in \textbf{Case 1} has the following solution
\begin{align}
{\textbf{Case 1:}} \ \ f_{\rm{Matrix}}(\bullet)&=
-{\rm{log}}|{\bf{Q}}^{\rm{H}}{\bf{F}}^{\rm{H}}{\bf{H}}^{\rm{H}}
{\bf{K}}_{\bf{F}}^{-1}{\bf{H}}{\bf{F}}{\bf{Q}}
+{\bf{N}}|  \nonumber \\
{\textbf{Solution:}} \ \ \ \ \ \  {\bf{Q}}_{\rm{opt}}&={\bf{U}}_{\bf{FHF}}{\bf{\bar U}}_{\bf{N}}^{\rm{H}}.
\end{align}

\noindent {\textbf{\textsl{Conclusion 2:}}} Based on (\ref{definitions}) and Inequality 3, it is proved in Appendix~\ref{App:Proof_Q} that ${\bf{Q}}_{\rm{opt}}$ in \textbf{Case 2} satisfies
\begin{align}
{\textbf{Case 2:}} \ \ f_{\rm{Matrix}}(\bullet) &=
-{\rm{log}}|{\bf{A}}^{\rm{H}}{\bf{Q}}^{\rm{H}}{\bf{F}}^{\rm{H}}{\bf{H}}^
{\rm{H}}{\bf{K}}_{\bf{F}}^{-1}{\bf{H}}{\bf{F}}{\bf{Q}}{\bf{A}}
+{\bf{I}}|  \nonumber \\
{\textbf{Solution:}} \ \  \ \ \ \ {\bf{Q}}_{\rm{opt}}&={\bf{U}}_{\bf{FHF}}{\bf{U}}_{\bf{A}}^{\rm{H}}.
\end{align}

\noindent {\textbf{\textsl{Conclusion 3:}}} Based on (\ref{definitions}) and Inequality 4, ${\bf{Q}}_{\rm{opt}}$ in \textbf{Case 3} has the following solution
\begin{align}
{\textbf{Case 3:}} \ \ f_{\rm{Matrix}}(\bullet)&=
{\rm{Tr}}[({\bf{Q}}^{\rm{H}}{\bf{F}}^{\rm{H}}{\bf{H}}^
{\rm{H}}{\bf{K}}_{\bf{F}}^{-1}{\bf{H}}{\bf{F}}{\bf{Q}}
+{\bf{N}})^{-1}]  \nonumber \\
{\textbf{Solution:}} \ \ \ \ \ \ \
{\bf{Q}}_{\rm{opt}}&={\bf{U}}_{\bf{FHF}}{\bf{\bar U}}_{\bf{N}}^{\rm{H}}.
\end{align}

\noindent {\textbf{\textsl{Conclusion 4:}}} Similar to Conclusion 3, ${\bf{Q}}_{\rm{opt}}$ in \textbf{Case 4} satisfies
\begin{align}
{\textbf{Case 4:}} \ \ f_{\rm{Matrix}}(\bullet)&=
{\rm{Tr}}[({\bf{Q}}^{\rm{H}}{\bf{F}}^{\rm{H}}{\bf{H}}^
{\rm{H}}{\bf{K}}_{\bf{F}}^{-1}{\bf{H}}{\bf{F}}{\bf{Q}}\otimes{\bf{M}}
+{\bf{N}}\otimes{\bf{M}})^{-1}]  \nonumber \\
{\textbf{Solution:}} \ \ \ \ \ \ \
{\bf{Q}}_{\rm{opt}}&={\bf{U}}_{\bf{FHF}}{\bf{\bar U}}_{\bf{N}}^{\rm{H}}.
\end{align}

\noindent {\textbf{\textsl{Conclusion 5:}}} Based on (\ref{definitions}) and Inequality 2, it is proved in Appendix~\ref{App:Proof_Q} that ${\bf{Q}}_{\rm{opt}}$ in \textbf{Case 5} has the following solution
\begin{align}
{\textbf{Case 5:}} \ \ f_{\rm{Matrix}}(\bullet) &=
{\rm{log}}|{\bf{A}}^{\rm{H}}({\bf{Q}}^{\rm{H}}{\bf{F}}^{\rm{H}}
{\bf{H}}^{\rm{H}}{\bf{K}}_{\bf{F}}^{-1}{\bf{H}}{\bf{F}}{\bf{Q}}
+{\bf{I}})^{-1}{\bf{A}}+{\bf{N}}|  \nonumber \\
{\textbf{Solution:}} \ \ \ \ \  \ {\bf{Q}}_{\rm{opt}}&={\bf{U}}_{\bf{FHF}}{\bf{U}}_{\bf{ANA}}^{\rm{H}}.
\end{align}

\noindent {\textbf{\textsl{Conclusion 6:}}} Based on (\ref{definitions}) and Inequality 4, it is proved in Appendix~\ref{App:Proof_Q} that the optimal unitary matrix in \textbf{Case 6} has the following solution
\begin{align}
{\textbf{Case 6:}} \ \  f_{\rm{Matrix}}(\bullet)&= {\rm{Tr}}[{\bf{A}}^{\rm{H}}({\bf{Q}}^{\rm{H}}{\bf{F}}^{\rm{H}}
{\bf{H}}^{\rm{H}}{\bf{K}}_{\bf{F}}^{-1}{\bf{H}}{\bf{F}}{\bf{Q}}
+{\bf{I}})^{-1}{\bf{A}}] \nonumber \\
{\textbf{Solution:}} \ \ \ \ \ \ \  {\bf{Q}}_{\rm{opt}}&= {\bf{U}}_{\bf{FHF}}{\bf{U}}_{\bf{A}}^{\rm{H}}.
\end{align}

%

\subsection{Majorization Theory Based Methods}

\subsubsection{Additive majorization theory based method} $ \ $

\noindent {\textbf{\textsl{Conclusion 7:}}} When the objective function is additive Schur-concave function of diagonal elements of $({\bf{Q}}^{\rm{H}}{\bf{F}}^{\rm{H}}{\bf{H}}^{\rm{H}}
{\bf{K}}_{\bf{F}}^{-1}{\bf{H}}{\bf{F}}{\bf{Q}}+{\bf{I}})^{-1}$, ${\bf{Q}}_{\rm{opt}}$ should diagonalize ${\bf{F}}^{\rm{H}}{\bf{H}}^{\rm{H}}
{\bf{K}}_{\bf{F}}^{-1}{\bf{H}}
{\bf{F}}$ \cite{Palomar2006} and then we have the following solution
\begin{align}
{\textbf{Case 7:}} \ \  f_{\rm{Matrix}}(\bullet)&=
{ f}_{\rm{A-Schur}}^{\rm{Concave}}({\bf{d}}[({\bf{Q}}^{\rm{H}}
{\bf{F}}^{\rm{H}}{\bf{H}}^{\rm{H}}{\bf{K}}_{\bf{F}}^{-1}{\bf{H}}
{\bf{F}}{\bf{Q}}+{\bf{I}})^{-1}])  \nonumber \\
{\textbf{Solution:}} \ \  \ \ \ \ \  {\bf{Q}}_{\rm{opt}}&={\bf{U}}_{\bf{FHF}}^{\rm{H}}.
\end{align}On the other hand, if the objective function is additive Schur-convex, ${\bf{Q}}_{\rm{opt}}$ is the unitary matrix such that $({\bf{Q}}^{\rm{H}}{\bf{F}}^{\rm{H}}{\bf{H}}^{\rm{H}}{\bf{K}}_{\bf{F}}^{-1}{\bf{H}}{\bf{F}}{\bf{Q}}+{\bf{I}})^{-1}$
has identical diagonal elements and a typical solution is discrete fourier transform (DFT) matrix \cite{Palomar2006}. Therefore we have the following conclusion
\begin{align}
{\textbf{Case 7:}} \ \ f_{\rm{Matrix}}(\bullet)&=
{ f}_{\rm{A-Schur}}^{\rm{Convex}}({\bf{d}}[({\bf{Q}}^{\rm{H}}{\bf{F}}^{\rm{H}}
{\bf{H}}^{\rm{H}}{\bf{K}}_{\bf{F}}^{-1}{\bf{H}}
{\bf{F}}{\bf{Q}}+{\bf{I}})^{-1}])  \nonumber \\
{\textbf{Solution:}} \ \ \ \ \ \ \ {\bf{Q}}_{\rm{opt}}&={\bf{U}}_{\bf{FHF}}{\bf{Q}}_{\rm{DFT}}^{\rm{H}}
\end{align} where ${\bf{Q}}_{\rm{DFT}}$ is the DFT matrix.

\subsubsection{Multiplicative majorization theory based method} $ \ $

\noindent {\textbf{\textsl{Conclusion 8:}}} When the objective function is multiplicative Schur-concave function of squared diagonal elements of Cholesky factorization matrix of $({\bf{Q}}^{\rm{H}}{\bf{F}}^{\rm{H}}{\bf{H}}^{\rm{H}}{\bf{K}}_{\bf{F}}^{-1}{\bf{H}}{\bf{F}}{\bf{Q}}+{\bf{I}})^{-1}$, ${\bf{Q}}_{\rm{opt}}$ should diagonalize ${\bf{F}}^{\rm{H}}{\bf{H}}^{\rm{H}}
{\bf{K}}_{\bf{F}}^{-1}{\bf{H}}
{\bf{F}}$ \cite{Amico2008}. Then we have the following solution
\begin{align}
{\textbf{Case 8:}} \ \ f_{\rm{Matrix}}(\bullet)&=
{ f}_{\rm{M-Schur}}^{\rm{Concave}}({\bf{d}}^2[{\bf{L}}]) \nonumber \\
{\rm{with}} \ \ & ({\bf{Q}}^{\rm{H}}{\bf{F}}^{\rm{H}}{\bf{H}}^{\rm{H}}{\bf{K}}_{\bf{F}}^{-1}
{\bf{H}}
{\bf{F}}{\bf{Q}}+{\bf{I}})^{-1}={\bf{L}}{\bf{L}}^{\rm{H}} \nonumber \\
{\textbf{Solution:}} \ \ \ \ \ \ \ {\bf{Q}}_{\rm{opt}}&={\bf{U}}_{\bf{FHF}}.
\end{align}On the other hand, when the objective function is multiplicative  Schur-convex,  ${\bf{Q}}_{\rm{opt}}$ should make ${\bf{L}}$ have identical diagonal elements \cite{Amico2008}. The optimal unitary matrix ${\bf{Q}}_{\rm{opt}}$ has the following solution
\begin{align}
{\textbf{Case 8:}} \ \ f_{\rm{Matrix}}(\bullet)&=
{ f}_{\rm{M-Schur}}^{\rm{Convex}}({\bf{d}}^2[{\bf{L}}]) \nonumber \\
{\rm{with}} \ \  & ({\bf{Q}}^{\rm{H}}{\bf{F}}^{\rm{H}}{\bf{H}}^{\rm{H}}{\bf{K}}_{\bf{F}}^{-1}{\bf{H}}
{\bf{F}}{\bf{Q}}+{\bf{I}})^{-1}={\bf{L}}{\bf{L}}^{\rm{H}} \nonumber \\
{\textbf{Solution:}} \ \ \ \ \ \ \ {\bf{Q}}_{\rm{opt}}&={\bf{U}}_{\bf{FHF}}{\bf{Q}}_{\bf{T}}^{\rm{H}}
\end{align} where ${\bf{Q}}_{\bf{T}}$ is the unitary matrix which makes ${\bf{L}}$ have identical diagonal elements. How to construct ${\bf{Q}}_{\bf{T}}$ has been well studied and the interested readers are referred to \cite{Palomar2006}.

\section{Extensions to Multiple Matrix-Variate Cases}
\label{Sect: Multiple_Matrix_Variable}

In the previous sections, we have investigated the MMOPs with only one matrix variable. However in
 some more general cases, the optimization problems possibly involves multiple matrix variables. One example is the transceiver design for multi-hop AF MIMO relaying networks \cite{XingJSAC2012}. In this section, we take a step further to investigate the more general case with multiple matrix variables, which can be formulated as
\begin{align}
\label{MM}
& \min_{{\bf{X}}_k} \ \ \  {f}_{\rm{Matrix}}\left({\bf{K}}_{{\bf{X}}_1}^{-1/2}
{\bf{H}}_1{\bf{X}}_1,\hdots,{\bf{K}}_{{\bf{X}}_K}^{-1/2}
{\bf{H}}_K{\bf{X}}_K\right) \nonumber \\
& \ {\rm{s.t.}} \ \ \ \ {\bf{K}}_{{\bf{X}}_k}={\rm{Tr}}({\bf{X}}_k{\bf{X}}_k^{\rm{H}}{\boldsymbol \Psi}_k){\boldsymbol \Sigma}_k+{\sigma}_{n_k}^2{\bf{I}} \nonumber \\
 & \ \ \ \ \ \ \ \ \ {\rm{Tr}}({\bf{X}}_k{\bf{X}}_k^{\rm{H}}) \le P_k.
\end{align} Following a similar logic as that for  the single matrix variable case and introducing a series of unitary matrices ${\bf{Q}}_k$'s, we define the following new variables
\begin{align}
{\bf{F}}_k={\bf{X}}_{k}{\bf{Q}}_k^{\rm{H}}
\end{align} and then the optimization problem in (\ref{MM}) is reformulated as
\begin{align}
\label{Opt_Multiple_Matrix}
& \min_{{\bf{F}}_k,{\bf{Q}}_k} \ \  {f}_{\rm{Matrix}}\left({\bf{K}}_{{\bf{F}}_1}^{-1/2}
{\bf{H}}_1{\bf{F}}_1{\bf{Q}}_1,\hdots,{\bf{K}}_{{\bf{F}}_K}^{-1/2}
{\bf{H}}_K{\bf{F}}_K{\bf{Q}}_K\right) \nonumber \\
& \ {\rm{s.t.}} \ \ \ \ {\bf{K}}_{{\bf{F}}_k}={\rm{Tr}}({\bf{F}}_k{\bf{F}}_k^{\rm{H}}{\boldsymbol \Psi}_k){\boldsymbol \Sigma}_k+{\sigma}_{n_k}^2{\bf{I}} \nonumber \\
 & \ \ \ \ \ \ \ \ \ {\rm{Tr}}({\bf{F}}_k{\bf{F}}_k^{\rm{H}}) \le P_k \nonumber \\
 & \ \ \ \ \ \ \ \ \ {\bf{Q}}_k^{\rm{H}}{\bf{Q}}_k={\bf{I}}.
\end{align}The optimization problem (\ref{Opt_Multiple_Matrix}) considered in our work is assumed to have the property that for the optimal unitary matrices ${\bf{Q}}_{k}$'s, the objective function satisfies
\begin{align}
&{f}_{\rm{Matrix}}\left({\bf{K}}_{{\bf{F}}_1}^{-1/2}
{\bf{H}}_1{\bf{F}}_1{\bf{Q}}_1,\hdots,{\bf{K}}_{{\bf{F}}_K}^{-1/2}
{\bf{H}}_K{\bf{F}}_K{\bf{Q}}_K\right)\nonumber \\
\ge&{f}_{\rm{Matrix}}\left({\bf{K}}_{{\bf{F}}_1}^{-1/2}
{\bf{H}}_1{\bf{F}}_1{\bf{Q}}_{1,{\rm{opt}}},\hdots,{\bf{K}}_{{\bf{F}}_K}^{-1/2}
{\bf{H}}_K{\bf{F}}_K{\bf{Q}}_{K,{\rm{opt}}}\right)\nonumber \\
=&{f}_{\rm{Vector}}\left[\{{\boldsymbol \lambda}({\bf{F}}_k^{\rm{H}}{\bf{H}}_k^{\rm{H}}
{\bf{K}}_{{\bf{F}}_k}^{-1}
{\bf{H}}_k{\bf{F}}_k)\}_{k=1}^K\right],
\end{align}where ${f}_{\rm{Vector}}(\bullet)$ is a decreasing vector function with respect to the following big column vector
\begin{align}
[{\boldsymbol \lambda}^{\rm{T}}({\bf{F}}_1^{\rm{H}}{\bf{H}}_1^{\rm{H}}
{\bf{K}}_{{\bf{F}}_1}^{-1}
{\bf{H}}_1{\bf{F}}_1),\cdots ,{\boldsymbol \lambda}^{\rm{T}}({\bf{F}}_K^{\rm{H}}{\bf{H}}_K^{\rm{H}}
{\bf{K}}_{{\bf{F}}_K}^{-1}
{\bf{H}}_K{\bf{F}}_K) ]^{\rm{T}}.
\end{align}
 As a result, the optimization problem (\ref{Opt_Multiple_Matrix}) can be transformed as
\begin{align}
\label{Opt_V_M}
& \min_{{\bf{F}}_k} \ \ \ {f}_{\rm{Vector}}\left[\{{\boldsymbol \lambda}({\bf{F}}_k^{\rm{H}}{\bf{H}}_k^{\rm{H}}
{\bf{K}}_{{\bf{F}}_k}^{-1}
{\bf{H}}_k{\bf{F}}_k)\}_{k=1}^K\right] \nonumber \\
& \ {\rm{s.t.}} \ \ \ \ {\bf{K}}_{{\bf{F}}_k}={\rm{Tr}}({\bf{F}}_k{\bf{F}}_k^{\rm{H}}{\boldsymbol \Psi}_k){\boldsymbol \Sigma}_k+{\sigma}_{n_k}^2{\bf{I}} \nonumber \\
 & \ \ \ \ \ \ \ \ \ {\rm{Tr}}({\bf{F}}_k{\bf{F}}_k^{\rm{H}}) \le P_k.
\end{align} Fixing ${\bf{F}}_k$'s for $k\not=l$, the resulting optimization problem with the variable ${\bf{F}}_l$ is exactly the same as \textbf{MVOP 2} which is equivalent to \textbf{MMOP 1}. Based on the fundamental multi-variate optimization theory \cite{Boyd04},
the optimal solution of ${\bf{F}}_l$ is also the optimal solution of the problem with fixed ${\bf{F}}_k$'s for $k\not=l$. It means for the optimal solution the optimization problem (\ref{Opt_V_M}) can be decoupled into $K$ subproblems each of which is the same as \textbf{MVOP 2}. Therefore, using Theorem 1, the optimal solution of ${\bf{F}}_k$ has the following structure.

\noindent \textbf{{{\textsl{Theorem 2:}}}} Introducing the following auxiliary variables
\begin{align}
\label{eta_f_k}
\eta_{f_k} & \triangleq {\rm{Tr}}({\bf{F}}_k{\bf{F}}_k^{\rm{H}}{\boldsymbol \Psi}_k)\alpha_k+\sigma_{n_k}^2 \ \ \text{with} \ \ \alpha_k=\lambda_{\min}({\boldsymbol \Sigma}_k) \\
{\bf{K}}_{{\boldsymbol \Psi}_k} & \triangleq {\frac{P_k\lambda_{\max}({\boldsymbol \Psi}_k)}{{P_k\lambda_{\max}({\boldsymbol \Psi}_k)\alpha_k+\sigma_{n_k}^2}}{\boldsymbol \Sigma}_k+\frac{\sigma_{n_k}^2}
{{P_k\lambda_{\max}({\boldsymbol \Psi}_k)}\alpha_k+\sigma_{n_k}^2}{\bf{I}}}
\end{align}and defining a unitary matrix ${\bf{V}}_{{\bf{H}}_k}$ and a rectangular diagonal matrix ${\boldsymbol \Lambda}_{{\boldsymbol{\Pi}}_k}$ based on the following SVD
\begin{align}
& {\bf{K}}_{{\boldsymbol \Psi}_k}^{-1/2}{\bf{H}}_k(\alpha_k P_k{\boldsymbol \Psi}_k+{\sigma}_{n_k}^2{\bf{I}})^{-1/2}=
{\bf{U}}_{{\boldsymbol{\Pi}}_k}{\boldsymbol \Lambda}_{{\boldsymbol{\Pi}}_k}
{\bf{V}}_{{\boldsymbol{\Pi}}_k}^{\rm{H}} \ \ \text{with} \ \  {\boldsymbol \Lambda}_{{\boldsymbol{\Pi}}_k} \searrow,
\end{align}when ${\boldsymbol \Psi}_k \propto {\bf{I}}$ or ${\boldsymbol \Sigma}_k \propto {\bf{I}}$, the optimal solution of the optimization problem (\ref{Opt_V_M}) has the following structure
\begin{align}
\label{conslusion_2}
&{\bf{F}}_{k,\rm{opt}}=\sqrt{\eta_{f_k}}(\alpha_k P_k{\boldsymbol
\Psi}_k+\sigma_{n_k}^2{\bf{I}})^{-1/2}
{\bf{V}}_{{\boldsymbol{\Pi}}_k}{\boldsymbol \Lambda}_{{\bf{F}}_k}{\bf{U}}_{{\rm{Arb}},k}^{\rm{H}}
\end{align}where ${\bf{U}}_{{\rm{Arb}},k}$ is an arbitrary unitary matrix with a suitable dimension, ${\boldsymbol \Lambda}_{{\bf{F}}_k}$ is a rectangular diagonal matrix and the scalar $\eta_{f_k}$  equals
\begin{align}
&\eta_{f_k}=P_k/{\rm{Tr}}[(\alpha_k P_k{\boldsymbol
\Psi}_k+\sigma_{n_k}^2{\bf{I}})^{-1}{\bf{V}}_{{\boldsymbol{\Pi}}_k}
{\boldsymbol{\Lambda}}_{{\bf{F}}_k}
{\boldsymbol{\Lambda}}_{{\bf{F}}_k}^{\rm{H}}{\bf{V}}_{{\boldsymbol{\Pi}}_k}^{\rm{H}}].
\end{align}

Similar to the case with single matrix variable, when ${\boldsymbol \Psi}_k \not\propto {\bf{I}}$ and ${\boldsymbol \Sigma}_k \not\propto {\bf{I}}$, (\ref{conslusion_2}) is also a meaningful suboptimal solution. The values of the optimal ${\bf{Q}}_k$'s should also be determined according to the specific formulation of the investigated optimization problems. To illustrate how to compute ${\bf{Q}}_k$'s, in the following two kinds of representative examples are discussed in detail, e.g., serial structure and parallel structure. In the serial structure, the terms ${\bf{K}}_{{\bf{F}}_k}^{-1/2}
{\bf{H}}_k{\bf{F}}_k$'s are multiplied together in the objective function, while in the parallel structure they are separated with each other in the objective function.

\subsection{Serial Structure}

The investigated serial structure corresponds to the transceiver designs for multi-hop AF MIMO relaying networks. Generally speaking, there also exit two kinds of transceiver designs, i.e., linear transceiver designs and nonlinear transceiver designs. Linear transceiver designs have lower complexity than nonlinear ones. On the other hand, for given modulation schemes nonlinear transceiver designs usually have better performance than their linear counterparts in terms of  bit error rate (BER).
Then for serial structure, two representative examples are shown in the following for the illustration of the computation of ${\bf{Q}}_k$'s.

\noindent \textbf{Robust linear transceiver design for multi-hop AF MIMO relaying networks}

The robust linear transceiver designs for multi-hop AF MIMO relaying networks with additive Schur-convex/concave objective functions can be formulated as \cite{XingTSP2013}
\begin{align}
{\textbf{Case 9:}} \ & \min_{{\bf{F}}_k,{\bf{Q}}_k} \ \  {f}_{\rm{A-Schur}} \left({\bf{d}}[{\bf{I}}-{\bf{Q}}_1^{\rm{H}}{\bf{M}}_1^{\rm{H}}
\cdots{\bf{Q}}_K^{\rm{H}}{\bf{M}}_K^{\rm{H}}{\bf{M}}_K{\bf{Q}}_K\cdots{\bf{M}}_1{\bf{Q}}_1]\right) \nonumber \\
& \  {\rm{s.t.}}\ \ \ \ {\bf{M}}_k=[({\bf{K}}_{{\bf{F}}_k}^{-1/2}{\bf{ H}}_k{\bf{F}}_k)({\bf{K}}_{{\bf{F}}_k}^{-1/2}{\bf{ H}}_k{\bf{F}}_k)^{\rm{H}}+{\bf{I}}]^{-1/2}({\bf{K}}_{{\bf{F}}_k}^{-1/2}{\bf{ H}}_k{\bf{F}}_k)  \nonumber \\
&\ \ \ \ \ \ \ \ \ {\rm{Tr}}({\bf{F}}_k{\bf{F}}_k^{\rm{H}}) \le P_k \ \  {\bf{K}}_{{\bf{F}}_k}={\rm{Tr}}({\bf{F}}_k{\bf{F}}_k^{\rm{H}}{\boldsymbol \Psi}_k){\boldsymbol \Sigma}_k+{\sigma}_{n_k}^2{\bf{I}} \nonumber \\
&\ \ \ \ \ \ \ \ \ {\bf{Q}}_k{\bf{Q}}_k^{\rm{H}}={\bf{I}}.
\end{align}


%

\noindent \textbf{Robust transceiver design for multi-hop AF MIMO relaying networks with THP/DFE}

In order to enhance the signal detection performance, nonlinear precoders and equalizers can be utilized at the source node and destination node respectively.
For both the cases that THP precoder is adopted at the source and the DFE equalizer is used at the destination, the robust transceiver designs with multiplicative Schur-convex/concave objective functions for multi-hop AF MIMO relaying networks can be written as \cite{XingJSAC2012}
\begin{align}
{\textbf{Case 10:}} \ \ & \min_{{\bf{F}}_k,{\bf{Q}}_k} \ \
{f}_{\rm{M-Schur}} \left({\bf{d}}^2[{\bf{L}}]\right) \nonumber \\
& \  {\rm{s.t.}}\ \ \ \ {\bf{I}}-{\bf{Q}}_1^{\rm{H}}{\bf{M}}_1^{\rm{H}}
\cdots{\bf{Q}}_K^{\rm{H}}{\bf{M}}_K^{\rm{H}}{\bf{M}}_K{\bf{Q}}_K\cdots{\bf{M}}_1{\bf{Q}}_1={\bf{L}}{\bf{L}}^{\rm{H}} \nonumber \\
&\ \ \ \ \ \ \ \ \ {\bf{M}}_k=[({\bf{K}}_{{\bf{F}}_k}^{-1/2}{\bf{ H}}_k{\bf{F}}_k)({\bf{K}}_{{\bf{F}}_k}^{-1/2}{\bf{ H}}_k{\bf{F}}_k)^{\rm{H}}+{\bf{I}}]^{-1/2}({\bf{K}}_{{\bf{F}}_k}^{-1/2}{\bf{ H}}_k{\bf{F}}_k)\nonumber \\
& \ \ \ \ \ \ \ \ \ {\rm{Tr}}({\bf{F}}_k{\bf{F}}_k^{\rm{H}}) \le P_k \ \  {\bf{K}}_{{\bf{F}}_k}={\rm{Tr}}({\bf{F}}_k{\bf{F}}_k^{\rm{H}}{\boldsymbol \Psi}_k){\boldsymbol \Sigma}_k+{\sigma}_{n_k}^2{\bf{I}} \nonumber \\
&\ \ \ \ \ \ \ \ \ {\bf{Q}}_k{\bf{Q}}_k^{\rm{H}}={\bf{I}}.
\end{align}
For the above optimization problems in \textbf{Case 9} and \textbf{10}, the derivations of the optimal ${\bf{Q}}_k$ rely on majorization theory. As discussed in \cite{XingTSP2013} and \cite{XingJSAC2012}, the key foundation of the derivation is the following theorem  \cite{XingTSP2013,XingJSAC2012}.

\noindent {\textbf{{\textsl{Theorem 3:}}}} For any $K$ complex rectangular matrices with compatible dimensions, the following inequality holds,
\begin{align}
&\prod_{i=1}^{k}\sigma_i({\boldsymbol A}_1\cdots{\boldsymbol A}_{K-1} {\boldsymbol A}_K) \le \prod_{i=1}^{k}\sigma_i({\boldsymbol A}_N)\sigma_i({\boldsymbol A}_{N-1})\cdots \sigma_i({\boldsymbol A}_1) \nonumber \\
&k=1,2,\cdots,{\rm{dim}}\{{\boldsymbol A}_1,\cdots,{\boldsymbol A}_{K-1}, {\boldsymbol A}_K\}
\end{align}where ${\rm{dim}}\{{\boldsymbol A}_1,\cdots,{\boldsymbol A}_{K-1}, {\boldsymbol A}_K\}$ denotes the minimal numbers of the columns and rows of ${\boldsymbol A}_1,\cdots,{\boldsymbol A}_{K-1}, {\boldsymbol A}_K$ and $\sigma_i({\bf{Z}})$ denotes the $i^{\rm{th}}$ largest singular value of ${\bf{Z}}$.

Based on \textbf{Theorem 3} and SVD of ${\bf{M}}_k={\bf{U}}_{{\bf{M}}_k}{\boldsymbol{\Lambda}}_{{\bf{M}}_k}{\bf{V}}_{{\bf{M}}_k}^{\rm{H}}$, we have the following conclusion.

\noindent \textbf{{\textsl{Conclusion 9:}}} For \textbf{Case 9} and \textbf{10}, the optimal unitary matrices ${\bf{Q}}_{k}$ have the following solutions \cite{XingJSAC2012,XingTSP2013}
\begin{align}
{\bf{Q}}_{k,{\rm{opt}}}={\bf{V}}_{{\bf{M}}_{k}}{\bf{U}}_{{\bf{M}}_{k-1}}^{\rm{H}}, \ \ \ \ k=2,\cdots,K.
\end{align} It should be pointed out that the optimal solution ${\bf{Q}}_{1,{\rm{opt}}}$ is a bit more complicated. For both the additive Schur-concave and multiplicative Schur-concave objective functions, ${\bf{Q}}_{1,{\rm{opt}}}$ makes ${\bf{Q}}_1^{\rm{H}}{\bf{M}}_1^{\rm{H}}
\cdots{\bf{M}}_1{\bf{Q}}_1$ diagonalized \cite{XingJSAC2012,XingTSP2013}. On the other hand, for the additive Schur-convex objective function, ${\bf{Q}}_{1,{\rm{opt}}}$ is the matrix which makes ${\bf{Q}}_1^{\rm{H}}{\bf{M}}_1^{\rm{H}}
\cdots{\bf{M}}_1{\bf{Q}}_1$ have the same diagonal elements \cite{XingTSP2013}, while for the multiplicative Schur-concave objective function, ${\bf{Q}}_{1,{\rm{opt}}}$ is the matrix which makes ${\bf{L}}$ have the same diagonal elements \cite{XingJSAC2012}.

%
\begin{figure}[!ht]
\centering
\includegraphics[width=.5\textwidth]{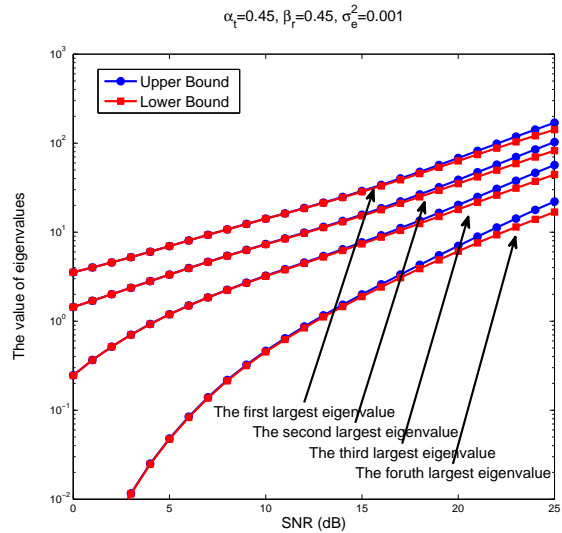}
\caption{The gaps between the upper bound and the lower bound with $\alpha_t=0.45$, $\beta_r=0.45$ and $\sigma_e^2=0.001$. }\label{fig:1}
\end{figure}
\subsection{Parallel Structure}
In contrast to the serial structure case, the optimization problem with parallel structure takes the sum of a series of matrix-monotone functions as its objective function, which is formulated as
\begin{align}
& \min_{{\bf{F}}_k,{\bf{Q}}_k} \ \ \  \sum_{k=1}^K{ f}_{{\rm{Matrix}},k}\left({\bf{Q}}_k^{\rm{H}}{\bf{F}}_k^{\rm{H}}
{\bf{H}}_k^{\rm{H}}{\bf{K}}_{{\bf{F}}_k}^{-1}
{\bf{H}}_k{\bf{F}}_k{\bf{Q}}_k\right) \nonumber \\
& \ \ {\rm{s.t.}} \  \ \ \  {\bf{K}}_{{\bf{F}}_k}={\rm{Tr}}({\bf{F}}_k{\bf{F}}_k^{\rm{H}}{\boldsymbol \Psi}_k){\boldsymbol \Sigma}_k+{\sigma}_{n_k}^2{\bf{I}} \nonumber \\
& \ \ \ \ \ \ \ \ \ \sum_k{\rm{Tr}}({\bf{F}}_k{\bf{F}}_k^{\rm{H}}) \le P ,
\end{align}where ${f}_{{\rm{Matrix}},i}(\bullet)$ is the matrix functions discussed in Section II. This kind of optimization problem corresponds to transceiver designs for MIMO-OFDM systems and can be decoupled into a series of subproblems which are exactly the ones discussed in the single matrix-variate case. Then it is straightforward that the same results as that given previously for single matrix-variate case can also be concluded.

\noindent \textbf{Remark:} For the more general case with multiple matrix variables, the derivation logic for computing the optimal diagonal matrices ${\boldsymbol \Lambda}_{{\bf{F}}_k}$ in (\ref{conslusion_2}) is exactly the same as that for the single matrix-variate case. In this paper, we only consider single user cases and our results cannot be simply extended to multi-user cases with mutual interference such as \cite{Baccarelli2003,Baccarelli2007}. This is a very interesting future research direction.

\section{Simulation Results and Discussions}
\label{Sect: Simulations}

As previously discussed, in the case of ${\boldsymbol{ \Psi}}{\not\propto} {\bf{I}}$ and ${\boldsymbol \Sigma}{\not\propto} {\bf{I}}$, the derived solution is only suboptimal, which maximizes a lower bound of the objective function of the original optimization problem (\ref{Matrix Optimization Problem}). A natural question is whether this bound is tight. In this section, some numerical results are used to assess the tightness of the lower bound. In order to show the tightness of the lower bound, an upper bound of the objective of (\ref{Matrix Optimization Problem}) is also given, which can be directly obtained by a simple replacement, i.e., replacing the equation $\alpha=\lambda_{\min}({{\boldsymbol \Sigma}})$ in (\ref{eta_f}) by $\alpha=\lambda_{\max}({{\boldsymbol \Sigma}})$. Then with this replacement, the optimization problem (\ref{MMOP3}) becomes to aim at maximizing an upper bound of the optimization problem (\ref{Matrix Optimization Problem}). It is interesting that the upper bound and lower bound have similar formulas. As the considered problems aiming to maximize an objective function, a lower bound is more meaningful than an upper bound.

In the following, we first consider a simple but representative model, e.g., a  point-to-point MIMO
system where both the transmitter and receiver are equipped with 4 antennas.
Furthermore, the channel errors are also taken into account. The estimation error
correlation matrices \begin{figure}
  \centering
  \subfigure[The case of $\alpha_t=0.9$, $\beta_r=0.3$ and $\sigma_e^2=0.01$.]
  {\label{fig:bound_1} 
    \includegraphics[width=2in]{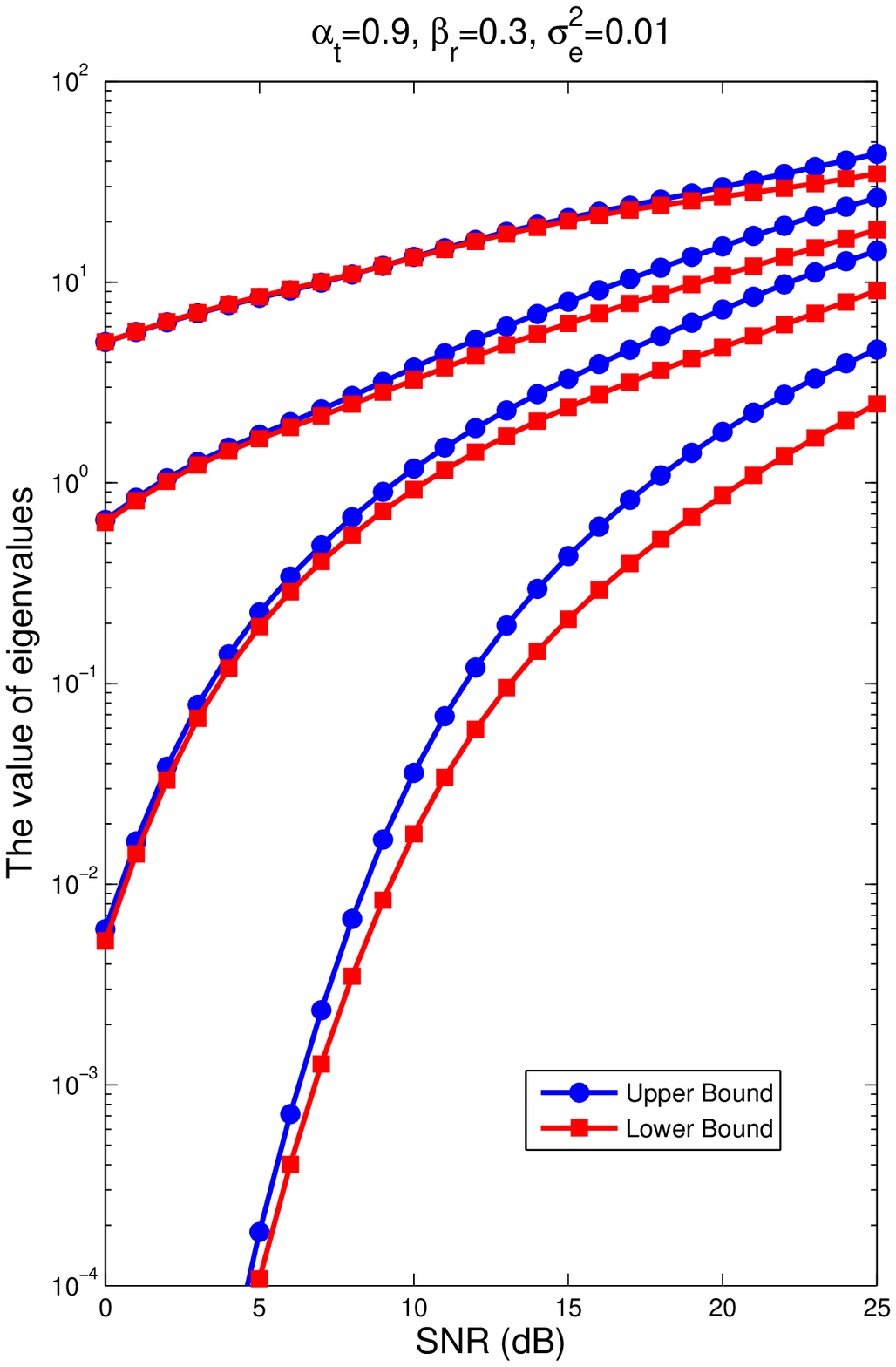}}
  \hspace{0.001in}
  \subfigure[The case of $\alpha_t=0.3$, $\beta_r=0.9$ and $\sigma_e^2=0.01$.]
  {
    \label{fig:bound_2} 
    \includegraphics[width=2in]{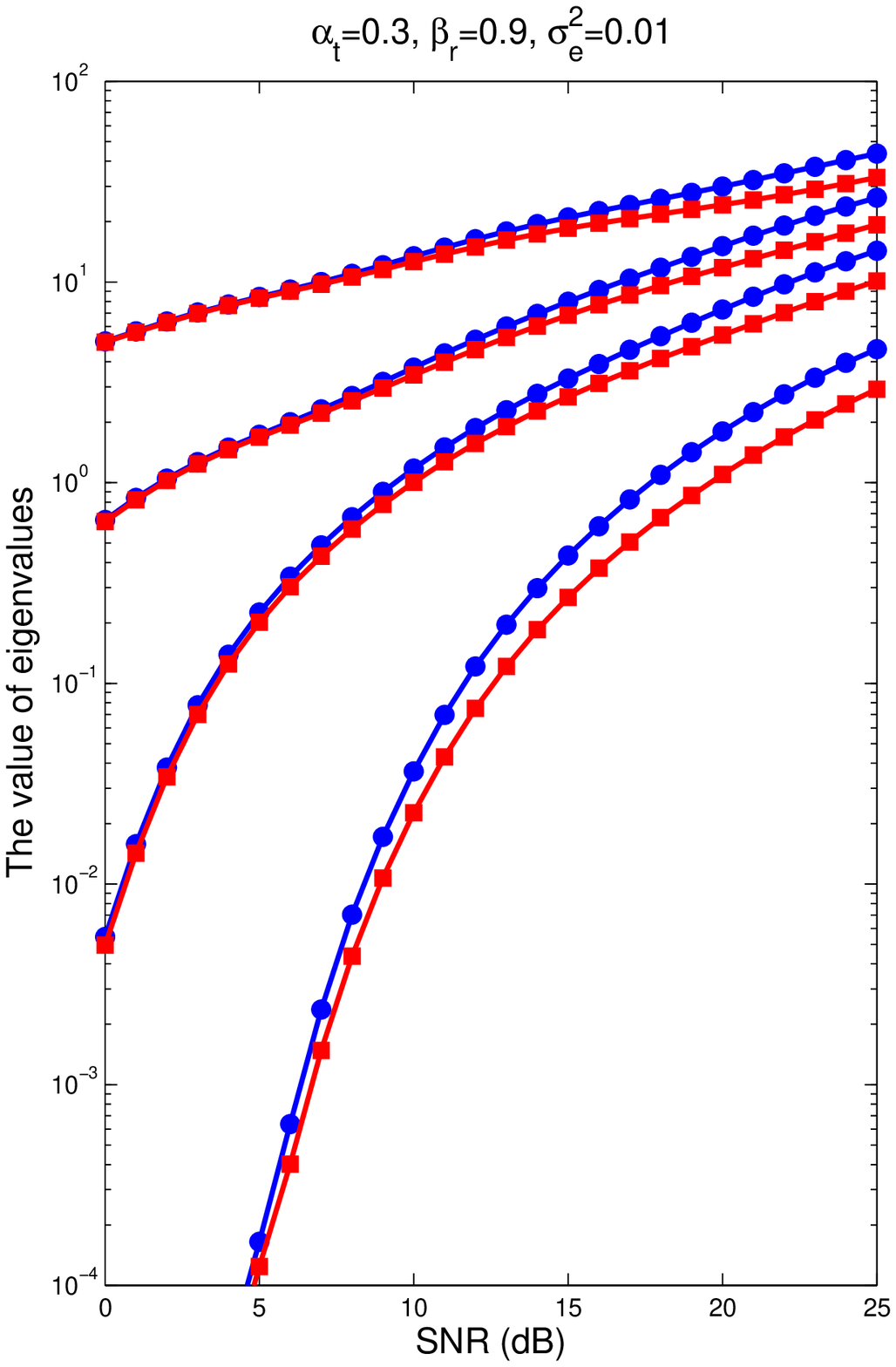}}
  \caption{The gaps between the upper and the lower bound for different settings.}
 \label{fig:bound} 
\end{figure}are chosen according to the popular exponential model  \cite{Xing10}, i.e.,
\begin{align}
 &\Delta{\bf{H}}\sim
\mathcal {C}\mathcal
{N}_{M,N}({\bf{0}}_{M,N},
{\boldsymbol
\Sigma}
\otimes {\boldsymbol \Psi}^{\rm{T}}),
\end{align}with
$[{\boldsymbol{\Psi}}]_{i,j}=\sigma_e^2\alpha_t^{|i-j|}$
and  $[{\boldsymbol{\Sigma}}]_{i,j}=\beta_r^{|i-j|}$, and $\sigma_{e}^2$ denotes
the estimation error variance. The estimated
channels ${\bf{\bar H}}$ is
generated based on the following complex Gaussian distributions \cite{Xing10}
\begin{align}
 &{\bf{\bar H}}\sim
\mathcal {C}\mathcal
{N}_{M,N}({\bf{0}}_{M,N},\frac{(1-\sigma_{e}^2)}{\sigma_{e}^2}
{\boldsymbol
\Sigma}
\otimes {\boldsymbol \Psi}^{\rm{T}}),
\end{align} such that channel realizations ${\bf{H}}={\bf{\bar
H}}+\Delta{\bf{H}}$ have unit variance. The signal-to-noise ratio (${\rm{SNR}}$) is defined as ${\rm{SNR}}=P/\sigma_{n}^2$.

As aforementioned, the transceiver design in this example in nature is to maximize a positive semi-definite matrix $ {\bf{\tilde F}}^{\rm{H}}{\boldsymbol \Pi}^{\rm{H}}{\boldsymbol \Pi}{\bf{\tilde F}}$. Specifically, it can be regarded as a multi-objective optimization problem with 4 objective functions (4 eigenvalues of  $ {\bf{\tilde F}}^{\rm{H}}{\boldsymbol \Pi}^{\rm{H}}{\boldsymbol \Pi}{\bf{\tilde F}}$). In the following, the 4 objective functions are plotted separately to show the gaps between the upper bound and lower bound.
In the following simulation figures, sum MSE is chosen as the objective function, which is one of most widely used performance metrics. As shown in Fig.~\ref{fig:1}, when the correlation factors at the transmitter and receiver are small and the estimation error is small, the gaps between the upper bound and lower bound are very small. In other words, in this kind of cases, the bound is much \begin{figure}
  \centering
  \subfigure[The case of $\alpha_t=0.9$ and $\beta_r=0.3$.]
  {\label{fig:MSE_1} 
    \includegraphics[width=2in]{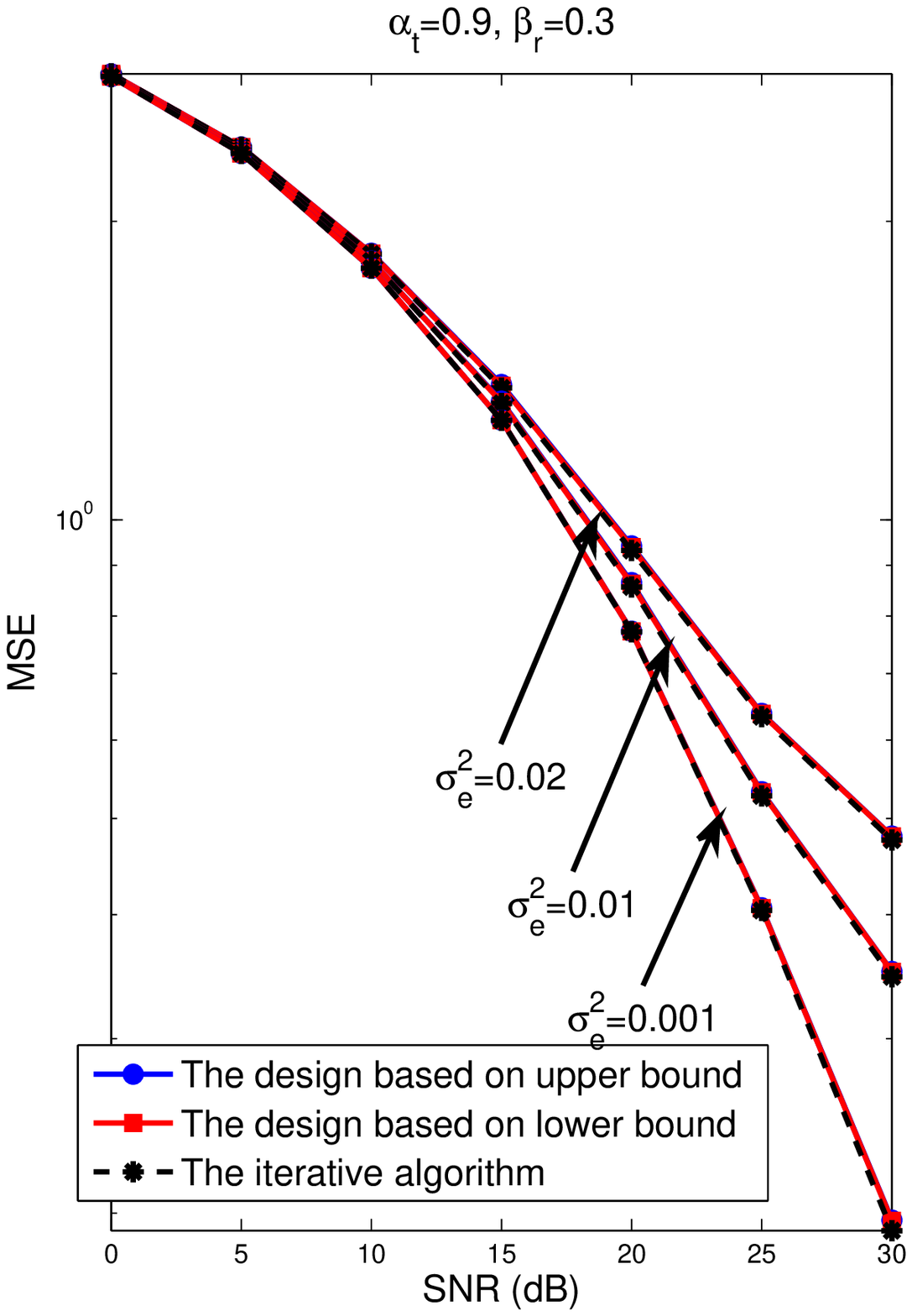}}
  \hspace{0.001in}
  \subfigure[The case of $\alpha_t=0.3$ and $\beta_r=0.9$.]
  {
    \label{fig:MSE_2} 
    \includegraphics[width=2in]{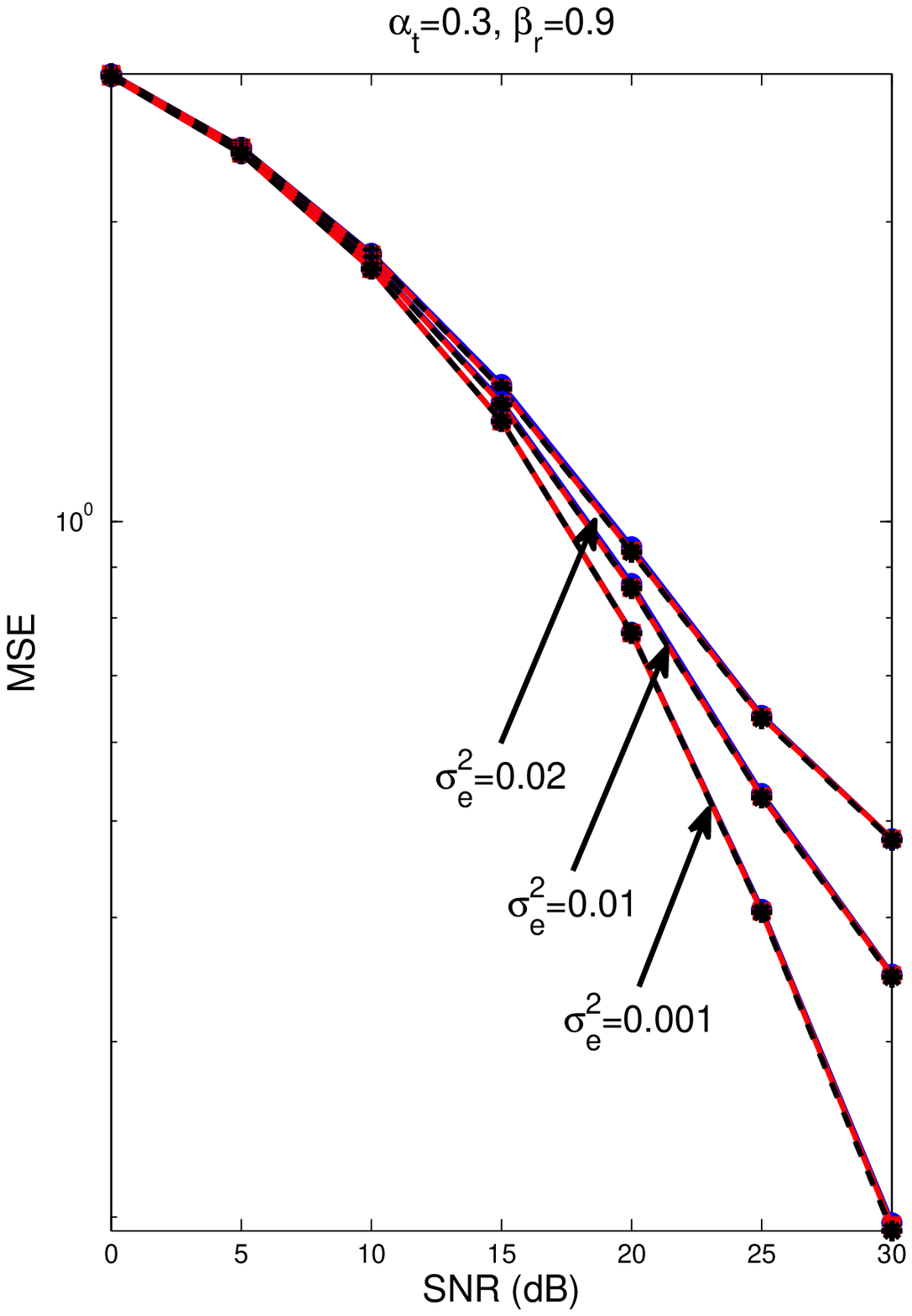}}
  \caption{The comparison of sum MSE among the design based on upper bound, the design based on lower bound and the iterative algorithm.}
 \label{fig:MSE} 
\end{figure}tight. For the case of ${\boldsymbol{ \Psi}}{\not\propto} {\bf{I}}$ and ${\boldsymbol \Sigma}{\not\propto} {\bf{I}}$, when the correlation factors ($\alpha_t$, $\beta_r$) and estimation error increase, the gaps will become slightly larger just as shown in Fig.~\ref{fig:bound}.

The curves shown in Figs.~\ref{fig:1} and ~\ref{fig:bound} are only the bounds instead of the real data detection MSEs. In other words, the curves shown in Figs.~\ref{fig:1} and \ref{fig:bound} are not the true values of the objective function of (\ref{Matrix Optimization Problem}). Then in Fig.~\ref{fig:MSE} we compare the performance of the solutions derived based on upper bound and lower bound in terms of the real data detection MSEs. The iterative algorithm proposed in \cite{Ding09} is also simulated as a benchmark. It can be seen that the three designs have exactly the same performance even with high correlation and large channel errors. It can be concluded that both the proposed solutions based on different bounds offer satisfied performance. As the solutions of the robust designs based on the upper and lower bounds have very similar structures, they have almost the same performance. It should be noted that compared to the iterative algorithm the proposed solutions have advantage in terms of computation complexity.

The complexity of computing the proposed closed-form solutions mainly comes from matrix inverse, matrix multiplication, matrix decomposition and water-filling. Notice that the complexity of water-filling is much lower than that of other matrix operations and then the dominated factors are matrix inverse, matrix multiplication and matrix decomposition. Note that for an $N\times N$ matrix the complexities of
these operations are all $\mathcal{O}(N^3)$. Then we simply count the numbers of these \begin{figure}
  \centering
  \subfigure[The case of $\alpha_{t,k}=0.6$ and $\beta_{r,k}=0.3$.]
  {\label{fig:AF_Capacity_1} 
    \includegraphics[width=2in]{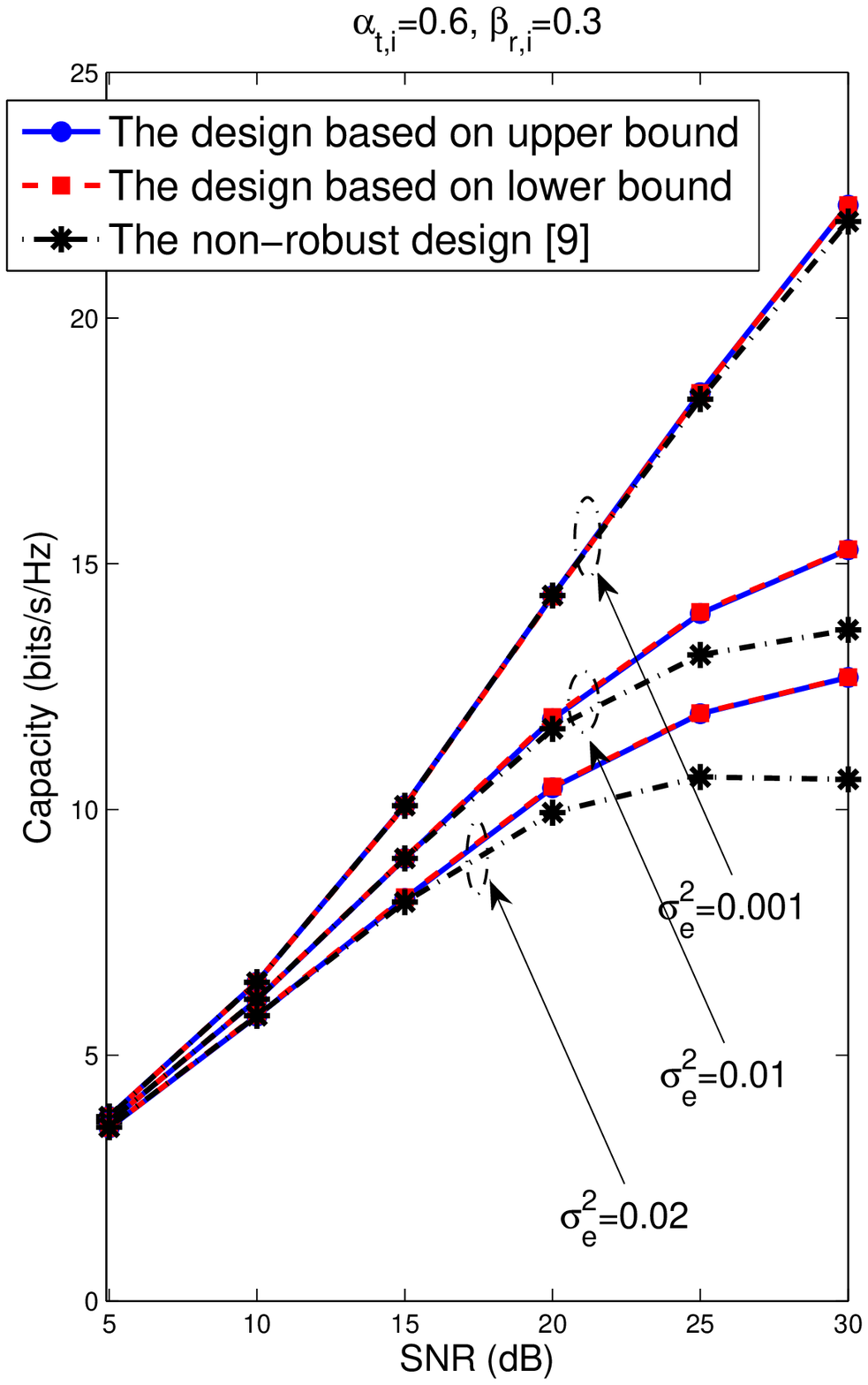}}
  \hspace{0.001in}
  \subfigure[The case of $\alpha_{t,k}=0.3$ and $\beta_{r,k}=0.6$.]
  {
    \label{fig:AF_Capacity_2} 
    \includegraphics[width=2in]{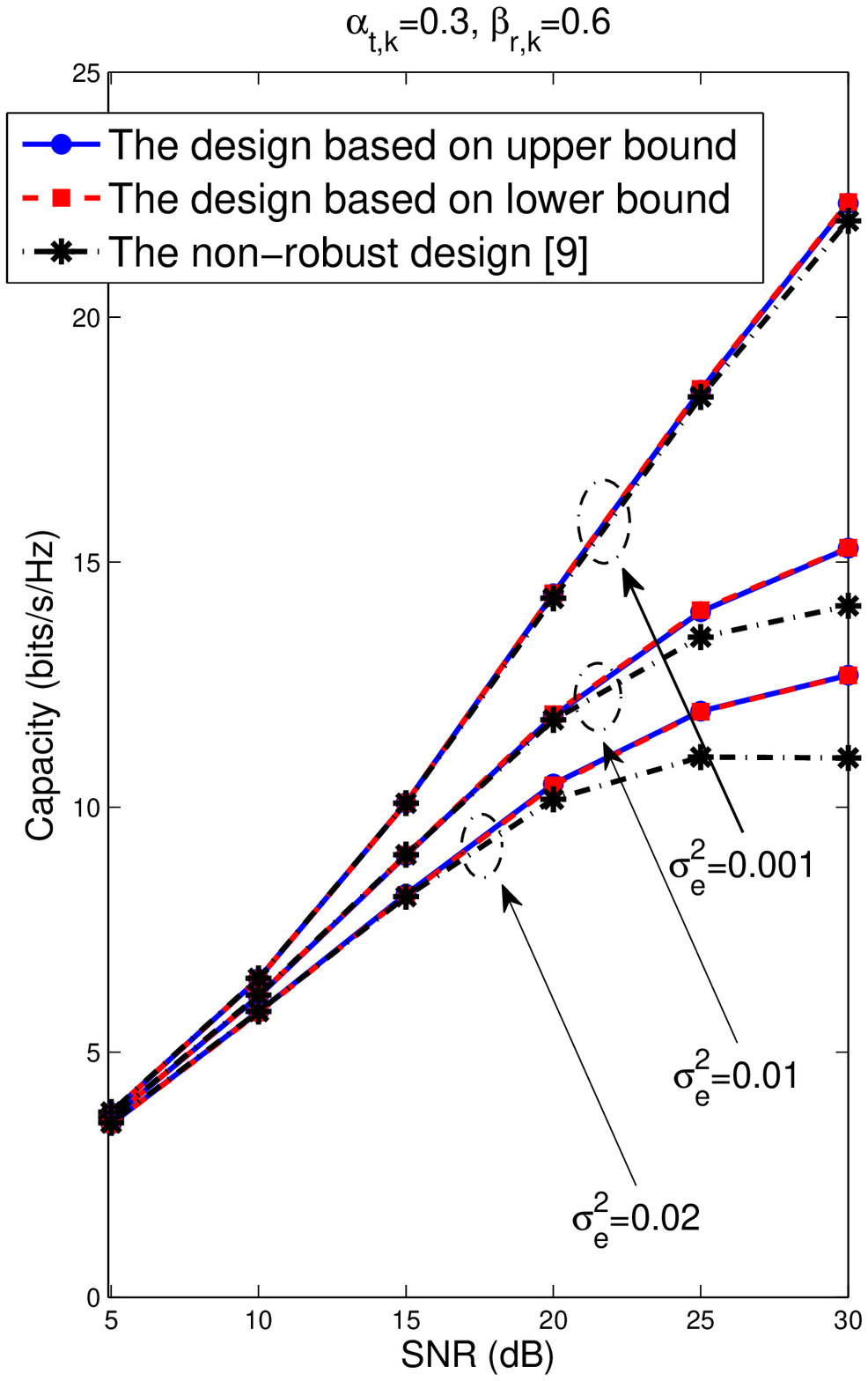}}
  \caption{The comparison of capacity among different designs for a three-hop AF MIMO relaying network.}
 \label{fig:AF MIMO Capacity} 
\end{figure}three operations to reflect computation complexity. For the proposed closed-form solution, the number is 11. With respect to the iterative algorithm, its complexity also comes from matrix inverse and matrix multiplication. Without doubt, the number of iterations must be taken into account in the complexity analysis. Unfortunately, it is very difficult to determine the number of iterations since it depends on initial values and many other factors. In the simulations it varies from 20 to 50 for different channel realizations. However, we discover that in each iteration the total number of matrix inverse and multiplication is 10. Then it is clear that our closed-form solution has an advantage in terms of complexity.

Taking a further step, a more general case with multiple matrix variables is also considered in this section. We investigate the linear transceiver design for three-hop AF MIMO relaying networks. In the considered network, all nodes are equipped with 4 antennas. The precoding matrix at the source and forwarding matrices at the two relay nodes are jointly designed. In each hop, similar to the above point-to-point MIMO system, the channel estimation errors are chosen as $\Delta{\bf{H}}_k\sim
\mathcal {C}\mathcal
{N}({\bf{0}},
{\boldsymbol
\Sigma}_k
\otimes {\boldsymbol \Psi}_k^{\rm{T}})$ and we also set $[{\boldsymbol{\Psi}}_k]_{i,j}=\sigma_{e}^2\alpha_{t,k}^{|i-j|}$
and  $[{\boldsymbol{\Sigma}}]_{i,j}=\beta_{r,k}^{|i-j|}$ \cite{XingTSP2013,XingJSAC2012}. For simplicity, it is assumed $\alpha_{t,1}=\alpha_{t,2}=\alpha_{t,3}$ and $\beta_{r,1}=\beta_{r,2}=\beta_{r,3}$ in our simulations. In addition,
\begin{figure}
  \centering
  \subfigure[The case of $\alpha_{t,k}=0.6$ and $\beta_{r,k}=0.3$.]
  {\label{fig:AF_MSE_1} 
    \includegraphics[width=2in]{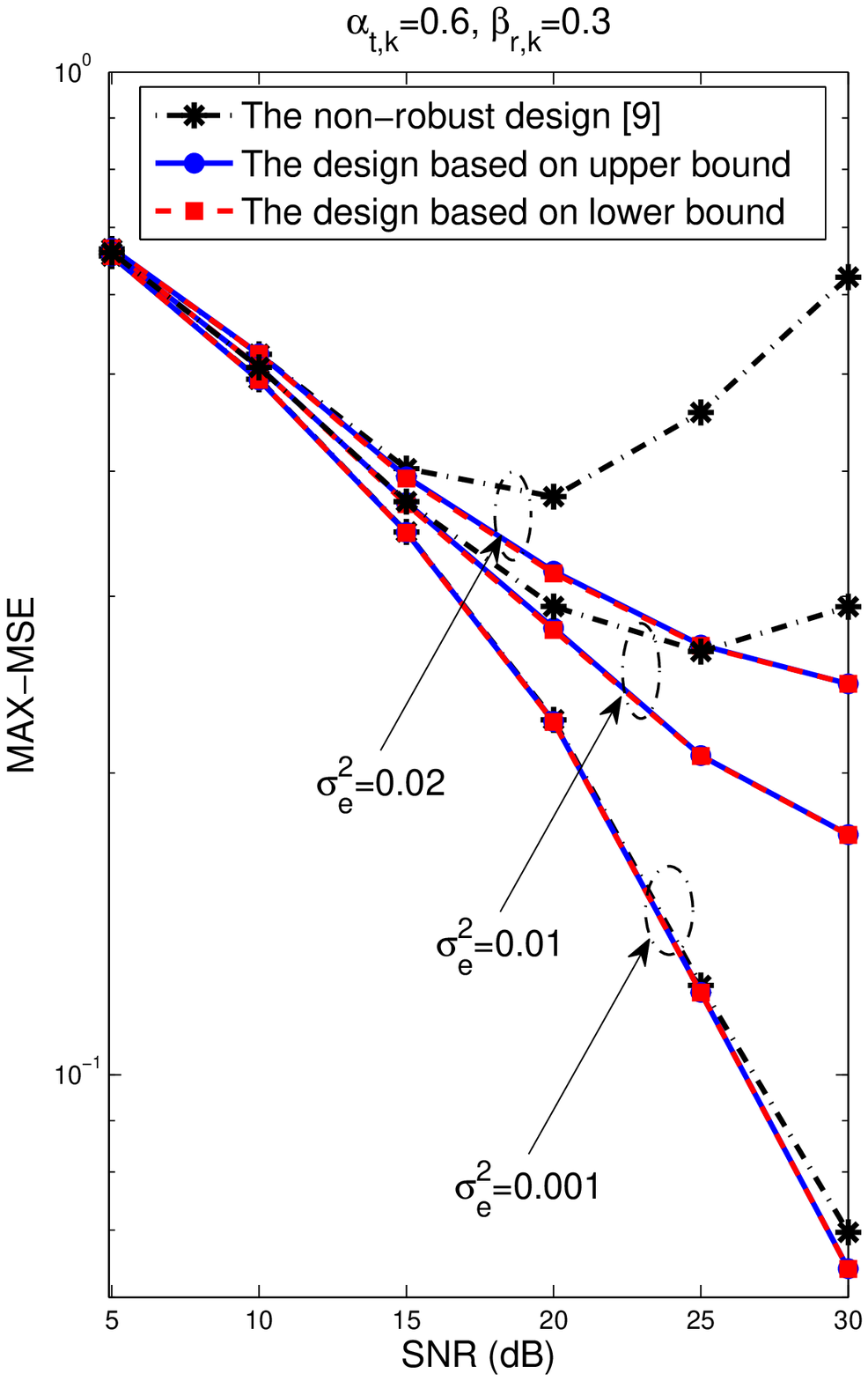}}
  \hspace{0.001in}
  \subfigure[The case of $\alpha_{t,k}=0.3$ and $\beta_{r,k}=0.6$.]
  {
    \label{fig:AF_MSE_2} 
    \includegraphics[width=2in]{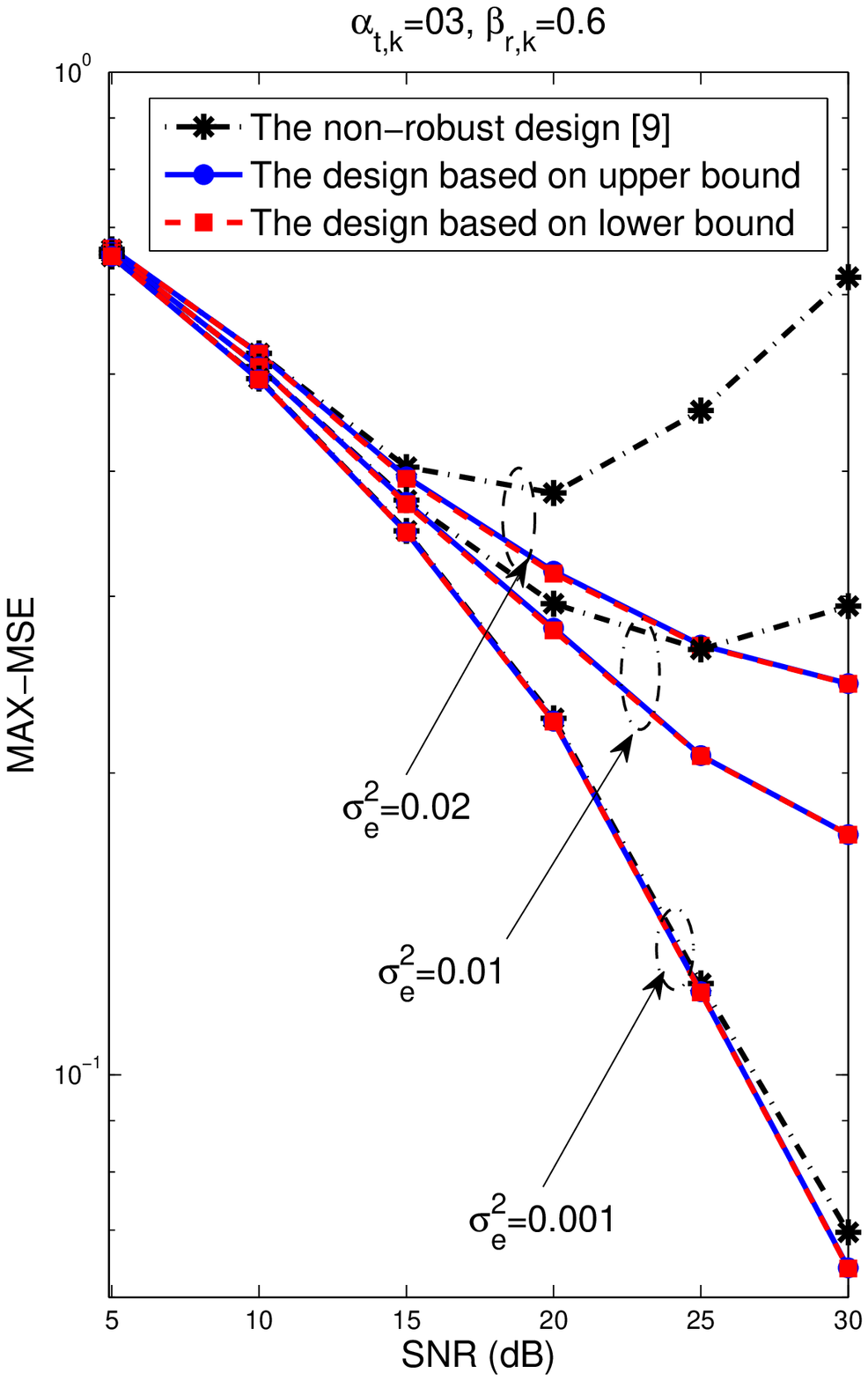}}
  \caption{The comparison of Max-MSE among different designs for a three-hop AF MIMO relaying network.}
 \label{fig:AF MIMO MSE} 
\end{figure}the channel estimates are also generated according to the matrix variate Gaussian distribution ${\bf{\bar H}}_k\sim
\mathcal {C}\mathcal
{N}({\bf{0}},\frac{(1-\sigma_{e}^2)}{\sigma_{e}^2}
{\boldsymbol
\Sigma}_k
\otimes {\boldsymbol \Psi}_k^{\rm{T}})$.  The SNR is defined as ${\rm{SNR}}=P_1/\sigma_{n_1}^2=P_2/\sigma_{n_2}^2=P_3/\sigma_{n_3}^2$.

Both additive Schur-concave and Shur-convex objective functions are chosen to assess the performance differences between the designs based on the upper and lower bounds for the case of ${\boldsymbol{ \Psi}}_k {\not\propto} {\bf{I}}$ and ${\boldsymbol \Sigma}_k {\not\propto} {\bf{I}}$. Maximizing the capacity of the three-hop AF MIMO relaying network is first investigated. For capacity maximization whose objective function is additive Shur-concave (it is also multiplicative Schur-concave), it is shown in Fig.~\ref{fig:AF MIMO Capacity} that performance gaps between the designs based the lower and upper bounds are negligible in the general case with different nonzero values of $\alpha_{t,k}$ and $\beta_{r,k}$. In other words, even in a complicated multi-hop AF MIMO relaying network, the robust designs based on the lower and upper bounds have almost the same performance, as their solutions have very similar structures. It can also be concluded that with channel estimation errors, both the robust designs based on lower and upper bounds have much better performance than the non-robust algorithm proposed in \cite{Rong2009TWC}, which regards the estimated CSI as the perfect CSI \cite{XingTSP2013,XingJSAC2012}. The simulation results demonstrate the effectiveness of the proposed transceiver structures.

Similar conclusions can also be drawn from Fig.~\ref{fig:AF MIMO MSE} for maximum MSE (MAX-MSE) minimization which has an additive Schur-convex objective function. For the optimization problem of MAX-MSE minimization, the design aims at minimizing the maximum diagonal element of the data detection MSE matrix at the destination instead of the sum. This design can realize a fairness between different data streams as it can always guarantee that the worst case can be optimized. It should be highlighted that different from Fig.~\ref{fig:MSE}, the curves in Fig.~\ref{fig:AF MIMO MSE} represent the maximum value of diagonal elements of the data MSE matrix instead of the sum value. For MAX-MSE minimization, the robust designs based on lower and upper bounds still have the same performance which is much better than that of the non-robust design proposed in \cite{Rong2009TWC} when channel errors are large.

From extensive simulation results, it is found that for different settings or other cases investigated in our work, similar results can always be achieved. Due to space limitation, those results are not shown in this paper. Here only the most representative cases are given.

\section{Conclusions}
\label{Sect: Conclusions}
In this paper, linear transceiver designs, nonlinear transceiver designs, training sequence designs and their robust counterparts for MIMO systems are addressed from a unified framework of matrix-monotonic optimization which takes advantage of the monotonicity in positive semi-definite matrices to simplify the considered optimization problems. It is discovered that exploiting the characteristics of matrix monotonicity, the optimal diagonalizable structures of unknown matrix variables can be exactly derived, which significantly simplify the investigated optimization problems. Furthermore, we also extend the optimization problems with single matrix variate to the more general ones with multiple matrix variates. This paper is the beginning of the research on matrix-monotonic optimization for wireless communications.
In nature monotonicity is a very fundamental characteristic which can generally be observed in wireless designs.
Therefore, it is expected that this elegant and powerful optimization framework can be applied to more wireless designs, e.g., transceiver designs or training designs for cognitive radio or physical layer secure communications. For future wireless communication systems such as 5G wireless communications, matrix-monotonic optimization will be very useful for Machine-to-Machine (M2M) communications which usually need multi-hop cooperative communications. In our future research, this matrix-monotonic optimization framework will be further developed for distributed MIMO networks with complicated mutual interference such as dense heterogeneous networks.

\section*{ACKNOWLEDGMENT}
The authors would like to thank the editor and anonymous reviewers for their professional and valuable comments/suggestions to improve the quality of this manuscript.


\appendices
%

\section{Proofs of \textbf{Lemmas 3} and \textbf{4}}
\label{App:Conclusion}

\noindent \textsl{Proof:} At the beginning we construct a new matrix ${{\boldsymbol{\tilde \Pi}}}$
\begin{align}
\label{equ_70}
{{\boldsymbol{\tilde \Pi}}}={\bf{U}}_{\boldsymbol{\Pi}}{\rm{diag}}\{[[{\Lambda}_{\boldsymbol{\Pi}}]_{1,1},\cdots, [{\Lambda}_{\boldsymbol{\Pi}}]_{N_{\boldsymbol{\Pi}},
N_{\boldsymbol{\Pi}}},\tau,\cdots,\tau]^{\rm{T}}\}{\bf{U}}_{\boldsymbol{\Pi}}^{\rm{H}}
\end{align}
where $N_{\boldsymbol{\Pi}}={\rm{Rank}}({\boldsymbol{\Pi}})$ and
$\tau$ is a real scalar which satisfies $0<\tau <[{\Lambda}_{\boldsymbol{\Pi}}]_{N_{\boldsymbol{\Pi}},N_{\boldsymbol{\Pi}}} $. Based on SVD, the second constraint is equivalent to the following equality \cite{Horn85}
\begin{align}
\label{F_F}
{\boldsymbol{{\Pi}}}{\bf{\tilde F}}=\sqrt{\rho} {\bf{U}}{\boldsymbol{{\Pi}}}{\bf{\tilde F}}_{\rm{in}}
\end{align}where ${\bf{U}}$ is an arbitrary unitary matrix. Multiplying both sides of the equality with ${\boldsymbol{{\tilde \Pi}}}^{-1}$, we directly have ${\boldsymbol{{\tilde \Pi}}}^{-1}{\boldsymbol{{\Pi}}}{\bf{\tilde F}}=\sqrt{\rho} {\boldsymbol{{\tilde \Pi}}}^{-1}{\bf{U}}{\boldsymbol{{\Pi}}}{\bf{\tilde F}}_{\rm{in}}$ based on which the scalar $\rho$ can be solved to be
\begin{align}
\label{app_rou}
\frac{{\rm{Tr}}({\boldsymbol{{\tilde \Pi}}}^{-1}{\boldsymbol{{\Pi}}}{\bf{\tilde F}}{\bf{\tilde F}}^{\rm{H}}{\boldsymbol{{\Pi}}}^{\rm{H}}{\boldsymbol{{\tilde \Pi}}}^{-1})}{{\rm{Tr}}({\boldsymbol{{\tilde \Pi}}}^{-1}
{\bf{U}}{\boldsymbol{{\Pi}}}{\bf{\tilde F}}_{\rm{in}}{\bf{\tilde F}}_{\rm{in}}^{\rm{H}}{\bf{U}}^{\rm{H}}{\boldsymbol{{\tilde \Pi}}}^{-1})
}=\rho.
\end{align}
Based on the definition of ${\boldsymbol{{\tilde \Pi}}}$ in (\ref{equ_70}), the numerator in (\ref{app_rou}) can be reformulated as
\begin{align}
\label{equ_73}
&{\rm{Tr}}({\boldsymbol{{\tilde \Pi}}}^{-1}{\boldsymbol{{\Pi}}}{\bf{\tilde F}}{\bf{\tilde F}}^{\rm{H}}{\boldsymbol{{\Pi}}}^{\rm{H}}{\boldsymbol{{\tilde \Pi}}}^{-1}) \nonumber \\
=&{\rm{Tr}}({\rm{diag}}[1,\cdots,1,0,\cdots,0]{\bf{V}}_{\boldsymbol{\Pi}}^{\rm{H}}{\bf{\tilde F}}{\bf{\tilde F}}^{\rm{H}}{\bf{V}}_{\boldsymbol{\Pi}}{\rm{diag}}
[1,\cdots,1,0,\cdots,0])\nonumber \\=&\sum_{i=1}^{N_{\boldsymbol{{\Pi}}}}
[{\bf{V}}_{\boldsymbol{\Pi}}^{\rm{H}}{\bf{\tilde F}}{\bf{\tilde F}}^{\rm{H}}{\bf{V}}_{\boldsymbol{\Pi}}]_{i,i}
\end{align}based on which we further have
\begin{align}
{\rm{Tr}}({\boldsymbol{{\tilde \Pi}}}^{-1}{\boldsymbol{{\Pi}}}{\bf{\tilde F}}{\bf{\tilde F}}^{\rm{H}}{\boldsymbol{{\Pi}}}^{\rm{H}}{\boldsymbol{{\tilde \Pi}}}^{-1})=\sum_{i=1}^{N_{\boldsymbol{{\Pi}}}}
[{\bf{V}}_{\boldsymbol{\Pi}}^{\rm{H}}{\bf{\tilde F}}{\bf{\tilde F}}^{\rm{H}}{\bf{V}}_{\boldsymbol{\Pi}}]_{i,i}
&\le \sum_{i=1}^{N_{\boldsymbol{{\Pi}}}}
\lambda_i({\bf{\tilde F}}{\bf{\tilde F}}^{\rm{H}})\le P. \label{inequality_1}
\end{align}
On the other hand, based on the matrix inequality $ {\sum}_i^N\lambda_i({\boldsymbol A})\lambda_{N-i+1}({\boldsymbol B})  \le {\rm{Tr}}({\boldsymbol A}{\boldsymbol B})$, the denominator in (\ref{app_rou}) satisfies
\begin{align}
& {\rm{Tr}}({\boldsymbol{{\tilde \Pi}}}^{-1}
{\bf{U}}{\boldsymbol{{\Pi}}}{\bf{\tilde F}}_{\rm{in}}{\bf{\tilde F}}_{\rm{in}}^{\rm{H}}{\boldsymbol{{\Pi}}}^{\rm{H}}{\bf{U}}^{\rm{H}}{\boldsymbol{{\tilde \Pi}}}^{-1})\ge
\sum_i \frac{\lambda_i({\boldsymbol{{\Pi}}}{\bf{\tilde F}}_{\rm{in}}{\bf{\tilde F}}_{\rm{in}}^{\rm{H}}{\boldsymbol{{\Pi}}}^{\rm{H}})}
{\lambda_i({\boldsymbol{{\tilde \Pi}}}^2)}.\label{inequality_2}
\end{align}Based on (\ref{inequality_1}) and (\ref{inequality_2}), together with the formulation of $\rho$ in (\ref{app_rou}), there exists an upper bound of $\rho$
\begin{align}
\rho\le \frac{P}{\sum_i \frac{\lambda_i({\boldsymbol{{\Pi}}}{\bf{\tilde F}}_{\rm{in}}{\bf{\tilde F}}_{\rm{in}}^{\rm{H}}{\boldsymbol{{\Pi}}}^{\rm{H}})}
{\lambda_i({\boldsymbol{{\tilde \Pi}}}^{2})}}
\end{align}where the equality holds if and only if both the equalities in (\ref{inequality_1}) and (\ref{inequality_2}) hold.

Before discussing how to achieve the upper bound, some unitary matrices are defined first based on the following SVDs
\begin{align}
{\boldsymbol{{\Pi}}}{\bf{\tilde F}}_{\rm{in}}&={\bf{U}}_{{\boldsymbol A}_{{\rm{in}}}}
{\boldsymbol \Lambda}_{{\boldsymbol A}_{{\rm{in}}}}{\bf{V}}_{{\boldsymbol A}_{{\rm{in}}}}^{\rm{H}} \ \ \text{with} \ \ {\boldsymbol \Lambda}_{{\boldsymbol A}_{{\rm{in}}}} \searrow \\
{\bf{\tilde F}}&={\bf{U}}_{{\bf{\tilde F}}}{\boldsymbol \Lambda}_{{\bf{\tilde F}}}{\bf{V}}_{{\bf{\tilde F}}}^{\rm{H}} \ \ \text{with} \ \ {\boldsymbol \Lambda}_{{\bf{\tilde F}}} \searrow.
\end{align}Note that as ${\rm{Tr}}({\bf{\tilde F}}{\bf{\tilde F}}^{\rm{H}})=P$, it is obvious that when ${\bf{U}}_{{\bf{\tilde F}}}={\bf{V}}_{\boldsymbol{\Pi}}$ and ${\rm{Rank}}\{{\bf{\tilde F}}\}\le N_{\boldsymbol{{\Pi}}} $ the numerator derived in (\ref{equ_73}) will achieve its maximum value
\begin{align}
\sum_{i=1}^{N_{\boldsymbol{{\Pi}}}}
[{\bf{V}}_{\boldsymbol{\Pi}}^{\rm{H}}{\bf{\tilde F}}{\bf{\tilde F}}^{\rm{H}}{\bf{V}}_{\boldsymbol{\Pi}}]_{i,i}=\sum_{i=1}^{N_{\boldsymbol{{\Pi}}}}
\lambda_i({\bf{\tilde F}}{\bf{\tilde F}}^{\rm{H}})=P.
\end{align}
On the other hand, the equality in (\ref{inequality_2}) holds when ${\bf{U}}={\bf{U}}_{{\boldsymbol{{\Pi}}}}{\bf{U}}_{{\boldsymbol A}_{{\rm{in}}}}^{\rm{H}}$. Substituting the two conclusions into (\ref{F_F}), it can be concluded that the optimal solutions have the following structure
\begin{align}
\label{Apped_X}
{\bf{\tilde F}}_{{\rm{opt}}}={\bf{V}}_{{\boldsymbol \Pi}}{\boldsymbol \Lambda}_{{\bf{F}}}{\bf{V}}_{{\boldsymbol A}_{{\rm{in}}}}^{\rm{H}} \ \ {\rm{with}} \ \ {\boldsymbol \Lambda}_{{\boldsymbol{\Pi}}}{\boldsymbol \Lambda}_{{\bf{F}}}={\boldsymbol \Lambda}_{{\boldsymbol A}_{{\rm{in}}}} \searrow.
\end{align} Notice that based on \textbf{Property 4}, the value of ${\bf{V}}_{{\boldsymbol A}_{{\rm{in}}}}$ does not affect the objective function and it can be an arbitrary unitary matrix.

\noindent $\blacksquare$

\section{Proofs of Matrix Inequalities}
\label{App:Proof_Inequ}
\noindent \textsl{Proof:} First, we investigate the left-hand side inequality which can be considered as the following optimization problem
\begin{align}
\label{App_U_opt}
& \min_{{\bf{Q}}} \ \ {\rm{Tr}}[({\bf{Q}}^{\rm{H}}{\boldsymbol {A}}{\bf{Q}}+{\boldsymbol B})^{-1}] \nonumber \\
& \ {\rm{s.t.}} \ \ \ \ {\bf{Q}}^{\rm{H}}{\bf{Q}}={\bf{I}}.
\end{align}The Lagrangian of (\ref{App_U_opt}) is \cite{Boyd04}
\begin{align}
{\mathcal {L}}={\rm{Tr}}[({\bf{Q}}^{\rm{H}}{\boldsymbol {A}}{\bf{Q}}+{\boldsymbol B})^{-1}]+{\rm{Tr}}[( {\bf{Q}}^{\rm{H}}{\bf{Q}}-{\bf{I}}){\boldsymbol \Phi}]
\end{align}where ${\boldsymbol \Phi}$ is the Lagrange multiplier, which is a Hermitian matrix. Then we can directly have the following KKT condition which is one of the necessary conditions for the optimal solutions
\begin{align}
& {\boldsymbol {A}}{\bf{Q}}({\bf{Q}}^{\rm{H}}{\boldsymbol {A}}{\bf{Q}}+{\boldsymbol B})^{-2}={\bf{Q}}{\boldsymbol \Phi} , \ \ \text{with} \ \ {\boldsymbol \Phi} = {\boldsymbol \Phi}^{\rm{H}}
\end{align}based on which and together with ${\bf{Q}}^{\rm{H}}{\bf{Q}}={\bf{I}}$ the following equality holds
\begin{align}
&{\bf{Q}}^{\rm{H}}{\boldsymbol {A}}{\bf{Q}}({\bf{Q}}^{\rm{H}}{\boldsymbol {A}}{\bf{Q}}+{\boldsymbol B})^{-2}={\boldsymbol \Phi}.
\end{align}As ${\bf{Q}}^{\rm{H}}{\boldsymbol {A}}{\bf{Q}}({\bf{Q}}^{\rm{H}}{\boldsymbol {A}}{\bf{Q}}+{\boldsymbol B})^{-2}$ is a Hermitian matrix, it can be concluded that ${\bf{Q}}^{\rm{H}}{\boldsymbol {A}}{\bf{Q}}$ and $({\bf{Q}}^{\rm{H}}{\boldsymbol {A}}{\bf{Q}}+{\boldsymbol B})^{-2}$ can be simultaneously diagonalized by a common unitary matrix \cite{Horn85}. In other words ${\bf{Q}}^{\rm{H}}{\boldsymbol {A}}{\bf{Q}}$ and ${\boldsymbol B}$ can be simultaneously diagonalized by a common unitary matrix. Then for the minimum value, the objective function equals
\begin{align}
& {\rm{Tr}}[({\bf{Q}}^{\rm{H}}{\boldsymbol {A}}{\bf{Q}}+{\boldsymbol B})^{-1}]=\sum_i\frac{1}{\lambda_i({\boldsymbol B})+{\widetilde{{ \lambda_i}({\boldsymbol A})}}}
\end{align}where $\{{\widetilde{{ \lambda_i}({\boldsymbol A})}}\}$ denotes a sequence consisting of the eigenvalues of ${\boldsymbol A}$ arranged in a certain order that should be determined. Following exactly the same logic, it can be proved that for the maximum value the objective function equals
\begin{align}
& {\rm{Tr}}[({\bf{Q}}^{\rm{H}}{\boldsymbol {A}}{\bf{Q}}+{\boldsymbol B})^{-1}]=\sum_i\frac{1}{\lambda_i({\boldsymbol B})+{\widehat{{{ \lambda_i}({\boldsymbol A})}}}}
\end{align}where $\{{\widehat{{{ \lambda_i}({\boldsymbol A})}}}\}$ also denotes a sequence of the eigenvalues of ${\boldsymbol A}$ in an arranged order. Then the problem of computing the maximum and minimum values becomes to determine the orders of eigenvalues of ${\boldsymbol A}$. For any two vectors with nonnegative elements, ${\bf{a}}$ and ${\bf{b}}$,  ${\bf{a}}_{\uparrow} + {\bf{b}}_{\downarrow}  \prec {\bf{a}}+{\bf{b}} \prec {\bf{a}}_{\uparrow} + {\bf{b}}_{\uparrow} $ where ${\bf{x}} \prec {\bf{y}}$ denotes vector ${\bf{x}}$ is majorized by vector ${\bf{y}}$ \cite{Marshall79}. In addition, ${\bf{x}}_{\uparrow}$ is the increasing rearrangement of ${\bf{x}}$ and ${\bf{x}}_{\downarrow}$ is the decreasing rearrangement of ${\bf{x}}$. Together with the fact that $1/x$ is a Schur-convex function, we directly have the following conclusion
\begin{align}{\sum_i}(\lambda_i({\boldsymbol {A}})+ \lambda_{N-i+1}({\boldsymbol {B}}))^{-1}\le {\rm{Tr}}[({\bf{Q}}^{\rm{H}}{\boldsymbol {A}}{\bf{Q}}+{\boldsymbol B})^{-1}]\le {\sum_i}(\lambda_i({\boldsymbol {A}})+ \lambda_i({\boldsymbol {B}}))^{-1},\end{align}where the left equality holds when
${\bf{Q}}={\bf{U}}_{{\boldsymbol {A}}}{\bf{\bar U}}_{{\boldsymbol {B}}}^{\rm{H}}$ and the right equality holds when
${\bf{Q}}={\bf{U}}_{{\boldsymbol {A}}}{\bf{ U}}_{{\boldsymbol {B}}}^{\rm{H}}$.

%

\noindent $\blacksquare$

\section{The Derivation of Optimal ${\bf{Q}}$}
\label{App:Proof_Q}
In this section, some complicated cases are investigated one by one.

\subsubsection{Case 2}Based on matrix inequality $|{\boldsymbol {A}}{\boldsymbol B}+{\bf{I}}|\le \prod_i(\lambda_i({\boldsymbol {A}}) \lambda_i({\boldsymbol {B}})+1)$, we have
\begin{align}
\label{case_d}
&{\rm{log}}|{\bf{A}}^{\rm{H}}{\bf{Q}}^{\rm{H}}{\bf{F}}^{\rm{H}}{\bf{H}}^{\rm{H}}{\bf{K}}_{\bf{F}}^{-1}
{\bf{H}}{\bf{F}}{\bf{Q}}{\bf{A}}+{\bf{I}}|\nonumber \\
=&{\rm{log}}|{\bf{Q}}^{\rm{H}}{\bf{F}}^{\rm{H}}{\bf{H}}^{\rm{H}}{\bf{K}}_{\bf{F}}^{-1}
{\bf{H}}{\bf{F}}{\bf{Q}}{\bf{A}}{\bf{A}}^{\rm{H}}+{\bf{I}}|\nonumber \\
\le &\sum_i{\rm{log}}[\lambda_{i}({\bf{A}}{\bf{A}}^{\rm{H}})
\lambda_i({\bf{F}}^{\rm{H}}{\bf{H}}^{\rm{H}}{\bf{K}}_{\bf{F}}^{-1}
{\bf{H}}{\bf{F}})+{\bf{I}}],
\end{align}where the first equality is based on the fact that $|{\boldsymbol {A}}{\boldsymbol {B}}+{\bf{I}}|=|{\boldsymbol {B}}{\boldsymbol {A}}+{\bf{I}}|$. Based on \textbf{Inequality 3}, the equality in (\ref{case_d}) holds when
\begin{align}
{\bf{Q}}={\bf{U}}_{{\bf{FHF}}}{\bf{U}}_{\bf{A}}^{\rm{H}}.
\end{align}

\subsubsection{Case 5} In Case 5, the objective function can be reformulated as
\begin{align}
&{\rm{log}}|{\bf{A}}^{\rm{H}}({\bf{Q}}^{\rm{H}}{\bf{F}}^{\rm{H}}{\bf{H}}^{\rm{H}}{\bf{K}}_{\bf{F}}^{-1}
{\bf{H}}{\bf{F}}{\bf{Q}}+{\bf{I}})^{-1}{\bf{A}}+{\bf{N}}|\nonumber \\
=&{\rm{log}}|{\bf{N}}|+{\rm{log}}|({\bf{Q}}^{\rm{H}}{\bf{F}}^{\rm{H}}
{\bf{H}}^{\rm{H}}{\bf{K}}_{\bf{F}}^{-1}
{\bf{H}}{\bf{F}}{\bf{Q}}+{\bf{I}})^{-1}{\bf{A}}{\bf{N}}^{-1}{\bf{A}}^{\rm{H}}
+{\bf{I}}| \nonumber \\
=&{\rm{log}}|{\bf{N}}|+{\rm{log}}|
{\bf{A}}{\bf{N}}^{-1}{\bf{A}}^{\rm{H}}+{\bf{Q}}^{\rm{H}}({\bf{F}}^{\rm{H}}{\bf{H}}^{\rm{H}}
{\bf{K}}_{\bf{F}}^{-1}
{\bf{H}}{\bf{F}}+{\bf{I}}){\bf{Q}}|\nonumber \\
&-{\rm{log}}|({\bf{F}}^{\rm{H}}{\bf{H}}^{\rm{H}}{\bf{K}}_{\bf{F}}^{-1}
{\bf{H}}{\bf{F}}+{\bf{I}})|,
\end{align}where the first equality is also based on  $|{\boldsymbol {A}}{\boldsymbol {B}}+{\bf{I}}|=|{\boldsymbol {B}}{\boldsymbol {A}}+{\bf{I}}|$.
Using the left inequality of (\ref{inequ_2}), we directly have
\begin{align}
\label{case_b}
& {\rm{log}}|{\bf{A}}^{\rm{H}}({\bf{Q}}^{\rm{H}}{\bf{F}}^{\rm{H}}{\bf{H}}^{\rm{H}}
{\bf{K}}_{\bf{F}}^{-1}
{\bf{H}}{\bf{F}}{\bf{Q}}+{\bf{I}})^{-1}{\bf{A}}+{\bf{N}}|
  \ge\nonumber \\&{\rm{log}}|{\bf{N}}|+ \sum_i{\rm{log}}[\lambda_i({\bf{A}}{\bf{N}}^{-1}{\bf{A}}^{\rm{H}})
+\lambda_i({\bf{F}}^{\rm{H}}{\bf{H}}^{\rm{H}}{\bf{K}}_{\bf{F}}^{-1}
{\bf{H}}{\bf{F}}+{\bf{I}})]\nonumber \\
&-\sum_i{\rm{log}}[\lambda_i({\bf{F}}^{\rm{H}}{\bf{H}}^{\rm{H}}{\bf{K}}_{\bf{F}}^{-1}
{\bf{H}}{\bf{F}}+{\bf{I}})].
\end{align}To attain the minimum value in (\ref{case_b}), based on \textbf{Inequality 2}
the following equation should hold
\begin{align}
{\bf{Q}}={\bf{U}}_{\bf{ANA}}{\bf{\bar U}}_{\bf{FHF}}^{\rm{H}}.
\end{align}

\subsubsection{Case 6}
In Case 6, based on the left inequality of (\ref{inequ_1}), we directly have
\begin{align}
\label{case_a}
&{\rm{Tr}}[{\bf{A}}^{\rm{H}}({\bf{Q}}^{\rm{H}}{\bf{F}}^{\rm{H}}{\bf{H}}^{\rm{H}}{\bf{K}}_{\bf{F}}^{-1}
{\bf{H}}{\bf{F}}{\bf{Q}}+{\bf{I}})^{-1}{\bf{A}}]\ge \sum_i^{N_X} \lambda_{i}({\bf{A}}{\bf{A}}^{\rm{H}}) \lambda_{N-i+1}(({\bf{F}}^{\rm{H}}{\bf{H}}^{\rm{H}}{\bf{K}}_{\bf{F}}^{-1}
{\bf{H}}{\bf{F}}+{\bf{I}})^{-1}).
\end{align}Based on \textbf{Inequality 1}, the equality in (\ref{case_a}) holds when the following equation is satisfied
\begin{align}
{\bf{Q}}={\bf{U}}_{\bf{FHF}}{\bf{U}}_{\bf{A}}^{\rm{H}}.
\end{align}

\end{document}